\documentclass[sigconf]{acmart}

\usepackage{flushend}
\usepackage{subfiles}
\usepackage{enumitem}
\usepackage{color, colortbl}
\usepackage{tikz}
\usepackage{caption} 
\usepackage{comment}

\usepackage{hyperref}       
\usepackage{url}            
\usepackage{booktabs}       
\usepackage{amsfonts}       
\usepackage{nicefrac}       
\usepackage{microtype}      
\usepackage{lipsum}
\usepackage{amsthm}
\usepackage[normalem]{ulem}
\usepackage{algorithm}
\usepackage{algpseudocode}
\usepackage{subcaption}
\usepackage{siunitx}        
\renewrobustcmd{\bfseries}{\fontseries{b}\selectfont}
\renewrobustcmd{\boldmath}{}
\newrobustcmd{\B}{\bfseries}

\definecolor{Gray}{gray}{0.9}
\newcommand*\circled[1]{\tikz[baseline=(char.base)]{
            \node[shape=circle,fill,inner sep=1pt] (char) {\textcolor{white}{#1}};}}

\newcommand{\arxiv}[1]{\textcolor{black}{#1}}
\newcolumntype{a}{>{\columncolor{Gray}}r}
\newcolumntype{b}{>{\columncolor{lightgray}}r}

\begin{document}

\title{DGAP: Efficient Dynamic Graph Analysis on Persistent Memory}
\author{Abdullah Al Raqibul Islam}
\affiliation{
	\institution{Computer Science Department, \\ University of North Carolina at Charlotte}
	\city{Charlotte}
    \state{NC}
    \country{USA}
}
\email{aislam6@uncc.edu}

\author{Dong Dai}
\affiliation{
	\institution{Computer Science Department,  \\ University of North Carolina at Charlotte}
	\city{Charlotte}
    \state{NC}
    \country{USA}
}
\email{ddai@uncc.edu}

\begin{abstract}
    Dynamic graphs, featuring continuously updated vertices and edges, have grown in importance for numerous real-world applications. To accommodate this, graph frameworks, particularly their internal data structures, must support both persistent graph updates and rapid graph analysis simultaneously, leading to complex designs to orchestrate `fast but volatile' and `persistent but slow' storage devices. Emerging persistent memory technologies, such as Optane DCPMM, offer a promising alternative to simplify the designs by providing data persistence, low latency, and high IOPS together. In light of this, we propose DGAP, a framework for efficient dynamic graph analysis on persistent memory. 
    Unlike traditional dynamic graph frameworks, which combine multiple graph data structures (e.g., edge list or adjacency list) to achieve the required performance, DGAP utilizes a single mutable Compressed Sparse Row (CSR) graph structure with new designs for persistent memory to construct the framework. Specifically, DGAP introduces a \textit{per-section edge log} to reduce write amplification on persistent memory; a \textit{per-thread undo log} to enable high-performance, crash-consistent rebalancing operations; and a data placement schema to minimize in-place updates on persistent memory. 
    Our extensive evaluation results demonstrate that DGAP can achieve up to $3.2\times$ better graph update performance and up to $3.77\times$ better graph analysis performance compared to state-of-the-art dynamic graph frameworks for persistent memory, such as XPGraph, LLAMA, and GraphOne.
\end{abstract}

\maketitle

\thispagestyle{plain}
\pagestyle{plain}

\section{Introduction}
\label{sec:intro}

The ability to ingest new graph data continuously and analyze the latest graphs efficiently is crucial for many real-world applications today. For instance, cellular network operators need to address traffic hotspots in their networks as they are generated and identified~\cite{iyer2015celliq}. A dynamic graph framework that can both \textit{persistently store new graph updates} and \textit{perform complex graph analysis on the latest graph} is essential for supporting such applications. However, constructing such a framework is fundamentally challenging. Existing storage devices like SSDs, hard disks, or DRAM either lack persistence (as in volatile DRAM) or offer low performance on graph analysis (as in SSDs or hard disks). To handle both operations, graph frameworks must manage various storage devices, design unique data structures for each, and find a balance between them, leading to intricate systems. For example, GraphOne persists the graphs updates on SSD in Edge List (EL), conducts graph analysis on DRAM using Adjacency List (AL), and continuously synchronizes data between the two~\cite{kumar2019graphone}.

Recently, a new set of non-volatile or persistent memory devices (PMs) have emerged, such as Intel Optane DC Persistent Memory~\cite{inteloptane}. These devices can be accessed in bytes via the memory bus with data persistence guarantees. Compared to DRAM, PMs offer data persistence and greater density (e.g., Optane's 512GB/dimm vs. DRAM's 64GB/dimm). Compared to block-based devices, PMs allow byte-level access using \verb|load| and \verb|store| instructions with significantly lower latency (e.g., $\sim$300 {ns} vs. $\sim$100 {ms}) and higher IOPS (e.g., $\sim$10M vs. $\sim$500K for random writes)~\cite{suzuki2015survey,qureshi2009enhancing,yang2013memristive,qureshi2009scalable}. These characteristics suggest a promising alternative for building dynamic graph frameworks: \textit{employ PMs to serve both graph updates and graph analysis} for persistence, speed, and capacity. This approach further avoids the cost of data movements and reduces the complexity of coordinating multiple data structures on different storage devices. 
Although Intel has discontinued Optane PMs due to business reasons, millions of these devices remain available, and various new non-volatile memory solutions continue to emerge. We contend that designing high-performance storage systems on persistent memory devices remains both economically practical and beneficial, as evidenced by recent studies~\cite{wang2022xpgraph,song2023prism}. 




However, directly porting existing graph frameworks to PMs can be sub-optimal. Existing dynamic graph frameworks, such as LLAMA~\cite{macko2015llama} or GraphOne~\cite{kumar2019graphone}, utilize block I/O interfaces, whose software overheads are not acceptable for byte-addressable PMs~\cite{wu2020lessons}. The data structures are not tailored for PMs either, leading to potential performance issues~\cite{islam2020performance,islam2022performance}. Moreover, although PMs are persistent devices, writing data persistently is complicated due to the existence of volatile CPU caches. 
Extra flushing and fencing operations, though necessary, become costly without the right optimizations~\cite{islam2020performance,islam2022performance}. Unexpected crashes further necessitates expensive transactions to avoid partial writes, significantly impacting the performance~\cite{haria2020mod, wu2021archtm}. 

On the other hand, existing PM-specific dynamic graph frameworks, such as NVGRAPH~\cite{lim2019enforcing} and, more recently, XPGraph~\cite{wang2022xpgraph}, continue to follow the traditional approach of coordinating separate persistence-friendly and analysis-friendly data structures (i.e., edge list or adjacency list) on DRAM or PMs. This approach still leads to overly complicated data synchronization between data structures and creates unnecessary conversions or movements. 

In this study, we introduce a novel approach to design a unified graph data structure, serving both graph persistence and analysis directly from persistent memory. 
To this end, we propose DGAP, a \underline{D}ynamic \underline{G}raph \underline{A}nalysis framework specifically designed for \underline{P}ersistent memory.  
DGAP is built upon a recently proposed mutable Compressed Sparse Row (CSR) graph structure~\cite{wheatman2021parallel,islam2022vcsr}, which leverages Packed Memory Array (PMA)~\cite{bender2007adaptive} for efficient graph updates and analysis. Instead of naively porting mutable CSR to PMs, DGAP introduces a series of new designs to enhance its performance on PMs. Firstly, DGAP introduces a new \textit{per-section edge log} data structure to mitigate the write amplification issues associated with mutable CSR. Secondly, DGAP integrates new \textit{per-thread undo logs} to support high-performance crash-consistent rebalancing operations, which are frequent and costly operations in mutable CSR. Thirdly, DGAP strategically caches various mutable CSR components in DRAM according to the workloads. Through these designs, DGAP is able to deliver exceptional performance on both graph updates and graph analysis by maximally utilizing PMs. 

We implemented DGAP in around 2,000 lines of C++ code and compared its performance to that of state-of-the-art graph frameworks on PMs, using multiple graph analysis algorithms on different real-world graphs. Our results show that DGAP achieves up to $3.2\times$ improved graph update performance and $3.77\times$ enhanced graph analysis performance compared to leading graph frameworks, such as XPGraph, LLAMA, and GraphOne.

{The remainder of this paper is organized as follows: In \S\ref{sec:background} we discuss the background and motivation of this study. We introduce persistent memory device, existing graph storage formats including PMA-based mutable CSR, and most importantly, why directly porting PMA-based mutable CSR to PMs does not work. In \S\ref{sec:design}, we present the key components of DGAP and its operations in details. We present the extensive experimental results in \S\ref{sec:eval}. We compare with related work in \S\ref{sec:related}, conclude this paper and discuss the future work in \S\ref{sec:conclusion}.}

\section{Background and Motivation}
\label{sec:background}


\subsection{PMs and Optane DCPMM Overview}
\label{subsec:optane}

\subsubsection{Overview}
Persistent memory describes storage devices that are accessible in bytes via memory interfaces and can retain the stored data after the power is off~\cite{lee2010phase,raoux2008phase,kultursay2013evaluating,akinaga2010resistive}. 
Intel Optane DC Persistent Memory is the first commercially available PMs~\cite{van2019persistent, inteloptane}. 
Working on Intel Cascade Lake platforms, Optane can scale up to 24TB in a single machine~\cite{cascadelake}. 
It can be configured in either \textit{Memory} mode or \textit{App Direct} mode~\cite{pmdk}. 
In \textit{Memory} mode, the Optane devices are exposed as DRAM, with the actual DRAM becomes a transparent `L4' cache to accelerate data access. However, this model does not support data persistence. In \textit{App Direct} mode, Optane devices are directly exposed to users alongside DRAM. This mode allows users to access both DCPMM and DRAM and offers data persistence capability. In this study, we focus on \textit{App Direct} mode. 

\subsubsection{Performance Features}
PMs exhibit performance characteristics critical for building graph storage on them. For instance, their writes are slower due to the added persistence cost. The performances of large sequential accesses are often better than small random accesses due to the internal read/write buffers in these devices. 
Here, we use Optane DCPMM as an example to further highlight some performance features~\cite{van2019persistent, izraelevitz2019basic,shu2018empirical,yang2020anempirical,xiang2022characterizing,islam2020performance,islam2022performance}.
{Firstly, the read/write performance of Optane DCPMM is asymmetric.} 
Write operations, particularly persistent ones, incur significant overheads (e.g., up to $\sim7$-$8\times$ slower than DRAM). In contrast, read latencies are around $\sim2$-$3\times$ slower than DRAM. 
This underscores the importance of minimizing unnecessary writes.
Secondly, since Optane DCPMM uses 256 bytes internal write buffers, small random writes will perform much worse than large sequential writes. It is then critical to ensure the writes can be properly grouped~\cite{linley2019first}. 

\subsubsection{Persistence Features}
The challenge to achieve persistence in PMs is that not all the components in the memory hierarchy is persistent. 
Optane DCPMM introduces a concept called Asynchronous DRAM Refresh (ADR) which ensures during a power loss, all data in ADR will be written to PMs. But
ADR does not include CPU caches.
To guarantee data persistence, programmers must explicitly call \texttt{CLFLUSHOPT} and {\texttt{SFENCE}} instructions to flush the cache line and enforce the memory operations order~\cite{chen2015persistent}. But even with the cache line flushed and memory fenced, large writes to PMs may still be partially persisted as its atomic write unit is small (i.e., 8 bytes). 
Transactions are essential for ensuring data safety during large writes, yet they can significantly affect the performance, as recent research suggested~\cite{haria2020mod, wu2021archtm}.
Lately, extended ADR (eADR) was introduced in the 3rd generation Intel Xeon Scalable Processors to make CPU caches included in the power fail protected domain~\cite{eadr}. The eADR feature greatly simplified the programming~\cite{zardoshti2020understanding}. But it is not available in all PMs platforms. The applications need to recognize the devices and perform correctly and efficiently regardless which platforms are supported. DGAP is implemented to work with both ADR and eADR platforms.

\subsection{Graph Store and CSR}

At the heart of graph frameworks are their storage data structures. There have been a significant number of graph storage data structures, such as edge list (EL), adjacency list (AL), Compressed Sparse Row (CSR), and many others~\cite{sahu2017ubiquity, besta2021practice} used in different graph frameworks~\cite{ediger2012stinger,shao2013trinity,kumar2019graphone,roy2013x,zhang2018graphit,kyrola2012graphchi,macko2015llama}.

EdgeList (EL) is a sequential edge log, efficient for edge additions but slow for vertex accesses since it requires scanning the entire edge log. The Adjacency List (AL) and its variations, like blocked adjacency list~\cite{pinar1999improving}, use a per-vertex linked list for storing vertex neighbors. While perform well at graph insertions and single vertex operations, they struggle with whole graph analysis due to memory overheads and cache inefficiencies~\cite{macko2015llama,kumar2019graphone}.

Compressed sparse row (CSR), on the other hand, is designed for efficient graph analysis. It groups all edges from the same vertex together and stores them sequentially in an edge array, while the vertex array stores each vertex's starting index. In this way, CSR supports both per-vertex queries and edge iterations efficiently. It delivers extreme graph analysis performance because most of the vertices and edges are accessed sequentially. Its major limitation, however, is that it can not accommodate dynamic graph updates without rebuilding the entire edge array for each edge insertion.
%
To address this limitation, recent studies have proposed to use the Packed Memory Array (PMA) to make the edge array mutable~\cite{sha2017technical, wheatman2018packed, deleo2021teseo, islam2022vcsr}. Such mutable CSR data structures can offer extreme graph analysis performance while handling graph updates efficiently, making them a perfect candidate to build the PMs-based graph framework. 

\subsection{PMA-based Mutable CSR}
\label{subsec:intro}
The {Packed Memory Array (PMA)} is fundamentally a sorted array with reserved empty gaps interspersed~\cite{bender2007adaptive}. These gaps provide room for future insertions without shifting the entire array. 
To maintain the gap density, PMA employs a binary \textit{PMA Tree} to track density changes in different sections of the array. For any section located at tree height $i$, PMA assigns the lower and upper bound density thresholds as $\rho_i$ and $\tau_i$. When insertions or deletions make the density of a section out of the range, PMA initiates \textit{rebalancing} operations to adjust its gap density by redistributing gaps among adjacent sections. The rebalancing will happen at a level where all affected sections' densities together will fall within the density range. If the whole array is full, PMA \textit{resizes} the array by increasing its size. The amortized write overhead for adaptive PMA is ${O}(\log{}N)$. More details about PMA can be found in~\cite{bender2007adaptive}.

PMA-based mutable CSRs incorporate this concept by replacing the original CSR edge array with the packed memory array, exemplified by PCSR~\cite{wheatman2018packed} first. 
VCSR~\cite{islam2022vcsr} further optimizes PCSR by considering the skewed workloads inherent in real-world graphs. It partitioned the edge array into varied-size sections and distributed the gaps unevenly based on historical workloads in each section to improve performance. 

\subsection{Issues of Mutable CSR on PMs}
\label{subsec:issues}
PMA-based mutable CSR has been proven effective to support both graph updates and analysis. However, due to the unique features of PMs, a naive implementation leads to problematic performance, as summarized in the later three issues. 


\subsubsection{Write Amplification Issue}
\begin{figure}[t]
	\centering
	\includegraphics[width=\linewidth]{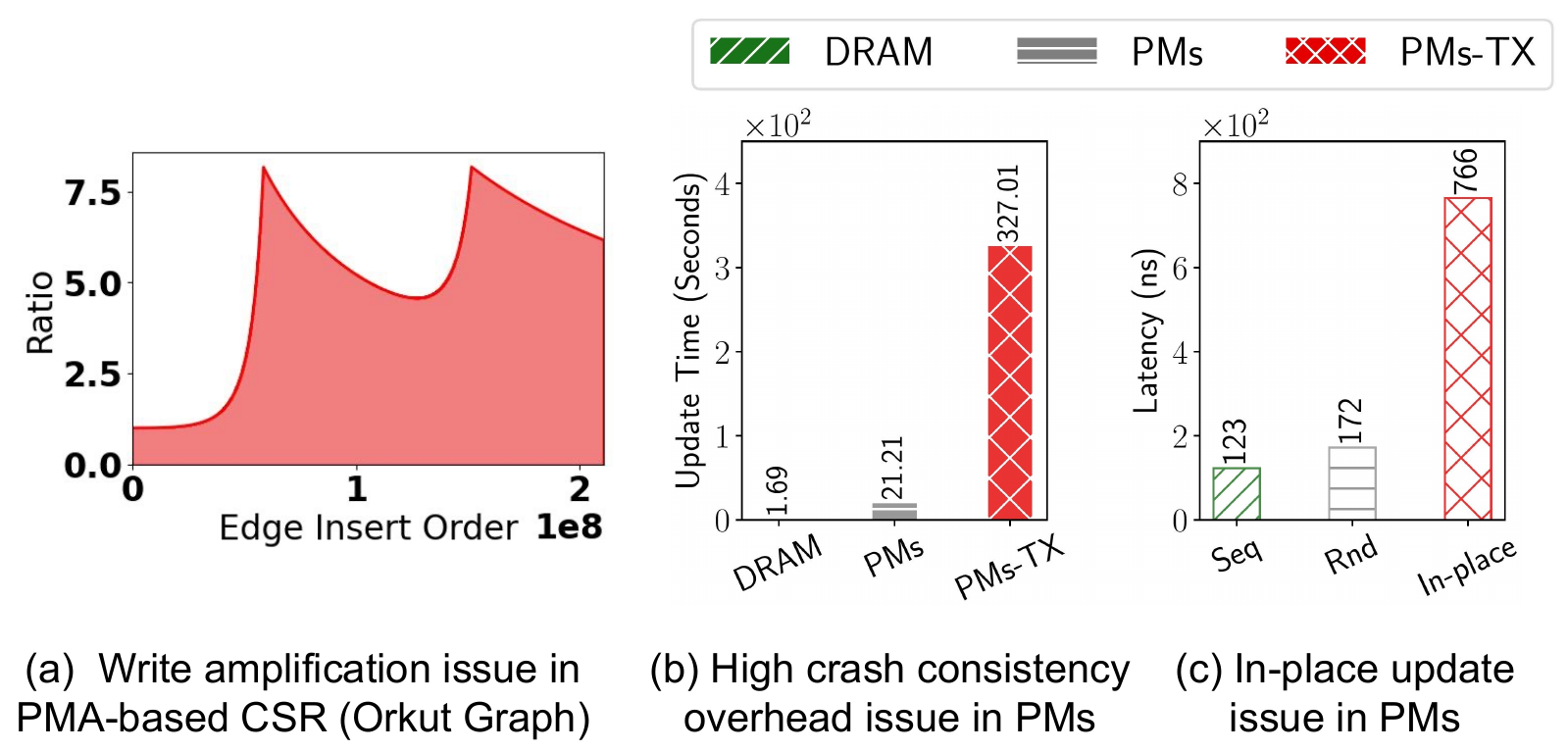}
    \caption[]{\small Issues of PMA-based mutable CSRs on PMs. \textit{The evaluation platform is described in Sec.~\S\ref{sec:eval:setup}}.}
	\label{motiv:vcsr_write_amp}
\end{figure}

Although mutable CSR avoids shifting the entire edge array for insertion, it still requires shifting a small range of elements if the targeted insertion location is occupied. These additional shifts result in write amplifications. Compared to DRAM, write amplifications in PMs are more critical due to PMs' asymmetric read/write performance. Additionally, these \textit{nearby shifts} often occur within a range smaller than 256 bytes, the size of the Optane DCPMM internal write buffer. This forces the buffers to be flushed before being filled, leading to inefficient buffer utilization. To illustrate the issue, we inserted the real-world graph, \textit{Orkut}~\cite{snapnets}, into a mutable CSR implementation and calculated the ratio of actual memory writes v.s the edge size (write amplification) during insertions. Figure~\ref{motiv:vcsr_write_amp}(a) reported the ratio during insertions. We can observe that the write amplification can be as high as $7\times$. It is hence critical to address such an issue. 



\subsubsection{Crash Consistency Issue.}
In addition to \textit{nearby shifts}, insertions could further trigger PMA \textit{rebalancing} when a section becomes full. These rebalancing operations move large chunks of sequential elements to new locations. 
Although efficient in DRAM, these operations are costly on PMs due to the persistence guarantee. 
It is necessary to use transactions to protect large chunks of writes. However, as demonstrated in Figure~\ref{motiv:vcsr_write_amp}(b), transactions are extremely expensive on PMs. The time required to insert a graph into DRAM, PMs (without transactions), and PMs-TX (with transactions) differ substantially~\cite{islam2020performance,islam2022performance}. Therefore, it is crucial to develop efficient crash recovery for frequent \textit{rebalancing} operations. 



\subsubsection{In-place Update Issue}
In-place updates in DRAM are efficient, leveraging the cache. But, persistent in-place updates on PMs are exactly the opposite.
Figure~\ref{motiv:vcsr_write_amp}(c) illustrates the performance of in-place updates on PMs. We present the latency of writing the same size of data in a sequential (Seq), random (Rnd), and in-place (In-place) manner respectively. We can observe $7\times$ difference in latency.
The reason is that persistent in-place updates repeatedly flush the same cache line and dramatically slow down the performance due to the blocking of previous flushing operations and possible on-chip wear-leveling protection~\cite{izraelevitz2019basic}. 
Crucial components of mutable CSR, such as the vertex degree and the PMA tree, require frequent in-place updates. Conducting these updates directly on PMs would be significantly slow.
It is essential to design the data placement strategy to minimize in-place updates on PMs.

\section{DGAP Design and Implementation}
\label{sec:design}

\begin{figure}[t]
	\centering
	\includegraphics[width=\linewidth]{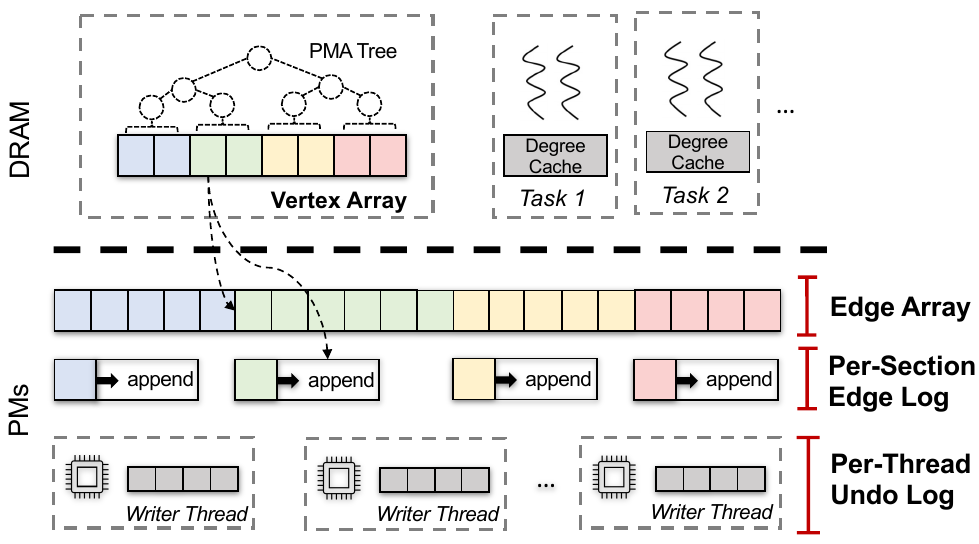}
	\caption{\small Overall architecture of DGAP.}
	\label{dgaparchi}
\end{figure}

DGAP, as illustrated in Fig.\ref{dgaparchi}, is designed to address the three issues outlined in Sec.~\S\ref{subsec:issues}. Its architecture comprises four primary components: \circled{1} \textit{vertex array}, \circled{2} \textit{edge array}, \circled{3} \textit{per-section edge log}, and \circled{4} \textit{per-thread undo log}. When interacting with DGAP, users launch multiple \textit{writer threads} for graph updates and can execute multi-thread graph analysis \textit{tasks} on the latest graphs. DGAP ensures the analysis tasks access only the latest graph snapshot when they start. This guarantees the long-running multi-iteration graph algorithms can access a consistent graph throughout their runs.

\subsubsection*{\circled{1} Vertex Array}
\label{sec:design:vertex}
DGAP stores all vertices sequentially in the vertex array. These sequential vertex IDs result from pre-processing by upstream applications, and their range is often known. Consequently, DGAP can pre-allocate the vertex array accordingly. Each vertex ($v$) in the vertex array takes 16 bytes to store three key pieces of metadata: its current degree ($degree_v$, 4 Bytes), starting index in the {edge array} ($start_v$, 8 Bytes), and a pointer to its \textit{per-section edge log} ($el_v$, 4 Bytes). 
The most important design decision about the DGAP vertex array is placing it entirely in DRAM. The main reason behind this design decision is to prevent frequent in-place updates on PMs. For dynamic graphs, the vertex degree ($degree_v$) must be updated each time an edge is inserted. The pointer to the \textit{per-section edge log} ($el_v$) also changes when edges are added to the edge log. Both operations are frequent enough to significantly impact overall performance if executed as in-place updates on PMs. Storing them in DRAM effectively avoids this issue.


Data safety is a critical issue when storing the entire vertex array in DRAM. DGAP introduces a new pivot element for each vertex in the {edge array} and leverages these elements to reconstruct the entire vertex array after a crash. More details are provided in Sec.~\S\ref{sec:operation:reboot}. The reconstruction is fast due to the high bandwidth of PMs for sequential accesses. Detailed results are reported in the evaluation section. 
Another potential concern is DRAM capacity. Theoretically, each DGAP vertex takes 16 bytes, so 16GB DRAM can store 1 billion vertices. Since most graphs have more edges than vertices, we anticipate that the capacity issue will primarily affect the PMs edge array rather than the DRAM vertex array.

\subsubsection*{\circled{2} Edge Array}
\label{sec:design:edge}
DGAP stores all the edges in the {edge array} on persistent memory. The edge array is a PMA constructed based on the VCSR strategy~\cite{islam2022vcsr}. Following other dynamic graph frameworks (e.g., XPGraph, GraphOne), each DGAP edge takes 4 bytes as it only stores the destination vertex ID. Storing the source vertex ID is unnecessary, as it is shared by all edges originating from the same vertex. The source vertex ID is instead stored as a pivot element at the beginning of each vertex' edge list. 
The pivot element serves as additional metadata in DGAP to reconstruct the DRAM vertex array after crashes. Specifically, the pivot is a special `edge' element with a value of \textminus\textit{vertex-id}. Since it is negative and illegal as a vertex ID, it can be used to denote the start of the vertex during recovery. Further details about DGAP recovery are in Sec.~\S\ref{sec:operation:reboot}.

One important design decision regarding the DGAP edge array is the storage order of all edges for a vertex. Traditionally, the edges of a vertex are sorted based on their destination vertex ID~\cite{wheatman2018packed}. However, DGAP stores them according to their insertion order, meaning a new edge will always be stored at the end of the vertex' edge list. So, an edge ($1 \to 2$) may be stored after edge ($1 \to 6$). This seemingly minor change is critical for DGAP to maintain a consistent snapshot of the latest graph for analysis tasks. This means that for any vertex $v$, if we know its degree at time $t$ ($degree_{v}^t$), we can easily determine its readable edges for $Task_t$, which should fall within the range [$start_{v}, start_{v}+degree_{v}^t$). Any edge after that will not be visible to $task_t$. Hence, creating a snapshot of the latest graph only involves storing the degrees of all vertices at time $t$. At present, we simply cache this degree info in the \textit{Degree Cache} within each task's DRAM space, as shown in Fig.~\ref{dgaparchi}. This can be done at the beginning of the analysis tasks. The primary issue here is memory cost. Many of the degrees are the same and do not need to be stored in each task. In the future, we plan to implement a Copy-on-Write (CoW) Degree Cache so that all tasks and the main vertex array can share unchanged degrees without wasting memory.

\subsubsection*{\circled{3} Per-section Edge Log}
\label{sec:design:log}
A primary performance challenge in existing PMA-based mutable CSRs on PMs is the write amplifications caused by \textit{nearby shifts} within each PMA section during insertions.
To mitigate this, our principal approach is to temporarily hold these insertions in a persistent log and merge them back in batches later. We introduce the concept of \textit{per-section edge logs} in DGAP, representing a pre-allocated, continuous, fixed-size space (\texttt{ELOG\_SZ}) on PMs dedicated for each PMA section. These logs temporarily store new edge insertions when a \textit{nearby shift} becomes necessary. In our prototype, \texttt{ELOG\_SZ} is set to {2K} bytes. 

Each element stored in the \textit{edge log} contains three metadata components and occupies {12} bytes: (i) source vertex ID, (ii) destination vertex ID, and (iii) a back-pointer. This back-pointer is designed to connect all edges originating from the same source vertex, arranging them in reverse order within the edge log. The most recent edge points back to the preceding edge of the same source vertex in the log. The edge log pointer ($el_v$), stored in the \textit{vertex array}, pinpoints the most recent edge of a vertex in the edge log. A detailed insertion workflow of DGAP is shown in Fig.~\ref{edgelog}.

When the \textit{per-section edge log} reaches 90\% usage, a \textit{merging} operation is initiated, integrating the edge log data back into the {edge array}. Notably, edges within the \textit{edge log} also contribute to the density of the corresponding \textit{edge array} section. Therefore, the standard PMA rebalancing operations might be triggered if either the edge array or edge log is approaching full capacity. During DGAP rebalancing, data from both the edge array sections and their respective edge logs are considered.
 


\subsubsection*{\circled{4} Per-thread Undo Log}
\label{sec:design:ulog}
PMA \textit{Rebalancing}, which redistributes gaps among sections, is critical for mutable CSR. To ensure data safety, it requires transaction mechanisms to avoid partial writes and guarantee crash consistency. 
While existing PMs programming libraries like PMDK~\cite{pmdk} support transactions natively, using them directly for recurrent \textit{rebalancing} operations results in significant overhead, due to two major bottlenecks: 1) the high memory allocation cost of frequent journal allocations and 2) performance overheads due to excessive ordering~\cite{haria2020mod}. 
In DGAP, we introduce a \textit{per-thread undo log} specifically to enhance the performance of \textit{rebalancing} while ensuring crash consistency. 
During insertion, whenever a \textit{Writer Thread} triggers rebalancing, before actually moving data, it first uses its own \textit{undo log} to persistently backup the data set to be relocated, chunk by chunk, acting as an `undo log'. If a crash happens in the middle, we can recover the data from the undo log. 
The \textit{per-thread undo log} is pre-allocated in fixed size (i.e., \texttt{ULOG\_SZ}) for each \textit{Writer Thread}. In our prototype, \texttt{ULOG\_SZ} is set to 2K bytes.

\subsection{DGAP Graph Operations}
\label{sec:operation}
This section explains how the DGAP components work together to serve various graph operations.

\subsubsection{Initialization}
When DGAP starts for the first time, it takes multiple user-specific parameters for system initialization. 
The number of vertices and edges in the graph are specified by the parameters \texttt{INIT\_VERTICES\_SIZE} and \texttt{INIT\_EDGES\_SIZE}. 
DGAP allocates the initial vertex array in DRAM and the edge array in PMs accordingly.
Both parameters are just initial user estimations. The actual numbers of vertices or edges can significantly surpass these values. When this happens, DGAP automatically resizes both the vertex and edge arrays during insertions. DGAP also utilizes the parameters \texttt{ELOG\_SIZE} and \texttt{ULOG\_SIZE} to pre-allocate the per-section edge logs and per-thread undo logs.
Furthermore, DGAP initializes multiple key metadata pieces on PMs for its operation. For instance, it maintains a global flag, \texttt{NORMAL\_SHUTDOWN}, on PMs to determine if DGAP had a graceful shutdown in its previous session. Whenever DGAP restarts, this value guides the system initialization process. 
In addition, DGAP creates and upholds various DRAM indexing metadata, including the PMA tree for density tracking. Locks are allocated based on this PMA tree to ensure concurrent reads/writes in DGAP. More details are discussed in later subsections.

\begin{figure}[t]
	\centering
	\includegraphics[width=0.9\linewidth]{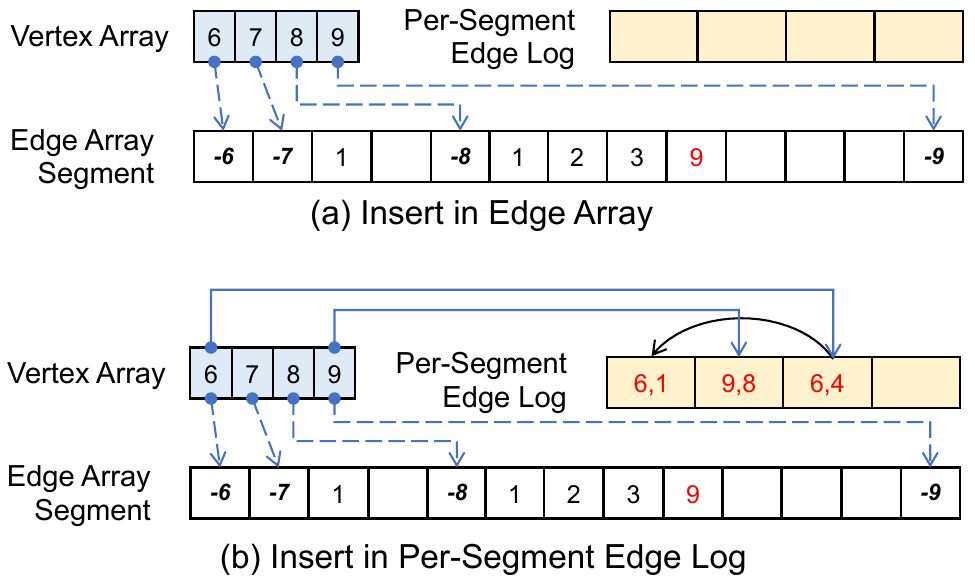}
	\caption{\small Two insertion cases in DGAP. \textit{The dashed blue line points to the starting index of a vertex in the \textit{edge array}.}}
	\label{edgelog}
\end{figure}

\subsubsection{Graph Updates}
\label{sec:operation:insertion}
DGAP utilizes the PMA-based mutable CSR structure to enable dynamic graph updates. For edge updates, an edge pair ($v_{src},v_{dst}$) will be fed to the \texttt{g.insertE()} call. For vertex updates, a vertex ID ($v_{src}$) will be fed to the \texttt{g.insertV()} call. Edge updates include both edge insertions and deletions. 
Deletions are executed by re-inserting the same edge marked with a tombstone flag. Specifically, we set the first bit of the destination vertex ID to 1, signifying that the edge has been removed from the graph. In the following, we delve into edge insertion operations.

Edge insertion includes two steps: 1) inserting the new edge into the \textit{edge array} or \textit{edge log}, and 2) updating the degree and pointer in the \textit{vertex array}. The DRAM vertex array is updated only after the PMs edge array has been successfully updated and flushed. In this way, even crash happens after PMs updates, the DRAM data structures can be reconstructed afterward. 
Given that we store all edges of a vertex chronologically, the insertion point for a new edge [$v_{src},v_{dst}$] in the \textit{edge array} can be easily determined. It can be calculated directly from the degree of $v_{src}$ and its starting index using the formula $(start_{v_{src}} + degree_{v_{src}})$. If the calculated location is a gap, the new edge can be inserted in an atomic manner. However, if the location is taken by a subsequent vertex, which requires a \textit{nearby shift}, DGAP appends the edge to the \textit{per-section edge log} to minimize write amplification.

Fig.~\ref{edgelog} illustrates two DGAP insertion scenarios. 
This figure provides a snapshot of the \textit{vertex array}, \textit{edge array}, and the associated \textit{per-section edge log}.
Here, edges for vertices ($6,7,8,9$) are showcased on the edge array with gaps, while the \textit{per-section edge log} is empty. Fig.~\ref{edgelog}(a) first shows a normal insertion case ($8\to9$) where the intended edge location is empty. Then the edge is inserted on the \textit{edge array} (marked in red). Fig.~\ref{edgelog}(b) shows another scenario where the desired locations for a series of edge insertions (e.g., $6\to1$, $6\to4$) are already taken (by vertex 7 and its edges). In this case, new edges will be stored on the \textit{per-section edge log} to reduce the unnecessary data shifts within the \textit{edge array}. Multiple edges of the same vertex on the edge log will be connected using the back-pointer, shown as the black arrow from ($6,4$) to ($6,1$) in Fig.~\ref{edgelog}(b). 

After many edge insertions, the corresponding section of the \textit{edge array} is becoming full. This will trigger a PMA rebalancing operation that redistributes the gaps among adjacent sections to ensure all the sections maintain a satisfactory density. 
While DGAP adopts the same logic to initiate the rebalancing, it carries out the operation with assistance from the \textit{per-thread undo log} to guarantee data consistency. Further details about crash-consistent rebalancing are elaborated in Sec~\ref{sec:operation:rebal}.

\subsubsection{Graph Analysis}
\label{sec:operation:read}

DGAP supports graph analysis by offering high-performance interfaces to iterate through all vertices (i.e., \texttt{g.v()}) and the edges associated with a vertex (i.e., \texttt{v.e()}). Graph analysis tasks might run for extended durations. For instance, the PageRank algorithm executed on the \textit{Orkut} graph can take over 20 seconds. During this time, the graph may be updated. To ensure a consistent view of the graph, it is necessary to guarantee that future reads from the same task bypass the newly added data.
To achieve this, users must first call the \texttt{g.consistent\_view()} function prior to iterating through the graph in their analysis tasks. Once this function is invoked, DGAP allocates a Degree Cache for the analysis task and temporarily holds the graph updates. It then copies the degree part of the vertex array to the per-task Degree Cache. This snapshot of degree information aids in pinpointing the appropriate set of edges for reading during task execution.

Once the snapshot is created, DGAP starts serving data-accessing function calls. For each call, DGAP initially reads the required metadata about $v$ from DRAM \textit{vertex array}, then accesses the PMs edge array based on that.
The necessary metadata from vertex array includes the starting index ($start_v$) and the edge log pointer ($el_v$). The degree information is obtained from the Degree Cache created at the task starting time $t$ ($degree_v^t$). If $el_v$ is \texttt{NULL}, then iterating through $v$'s edges involves simply iterating the corresponding \textit{edge array} from $start_v$ to ($start_v + degree_v^t$). If $el_v$ is not \texttt{NULL}, the edges also come from the edge log. In this case, we first scan the edge array. If the edge array does not contain a sufficient number of edges as needed (based on $degree_v^t$), DGAP proceeds to scan the edge log. The $el_v$ pointer always points to the last edge. From this point, we track all edges in the edge log through their back-pointer. To read only the required number of edges (assuming $rest_v^t$), we allocate a first-come-first-out (FIFO) buffer with a size of $rest_v^t$ to keep only the necessary edges.

\subsubsection{Crash Consistent PMA Rebalancing}
\label{sec:operation:rebal}

Thus far, we have discussed how DGAP handles point insertions without initiating rebalancing operations. 
In PMA, however, rebalancing is crucial when the insertions of new edges makes sections of the edge array overly dense. 
These rebalancing operations redistribute gaps among neighboring sections to alleviate the density issue.
Given that rebalancing involves considerable amount of data movement, it necessitates crash-consistent transactions. Yet, standard PMDK transactions are proven to be overly expensive. As a solution, DGAP introduces \textit{per-thread undo logs} to achieve more efficient, crash-consistent DGAP rebalancing.

\begin{figure}[t]
	\centering
	\includegraphics[width=0.9\linewidth]{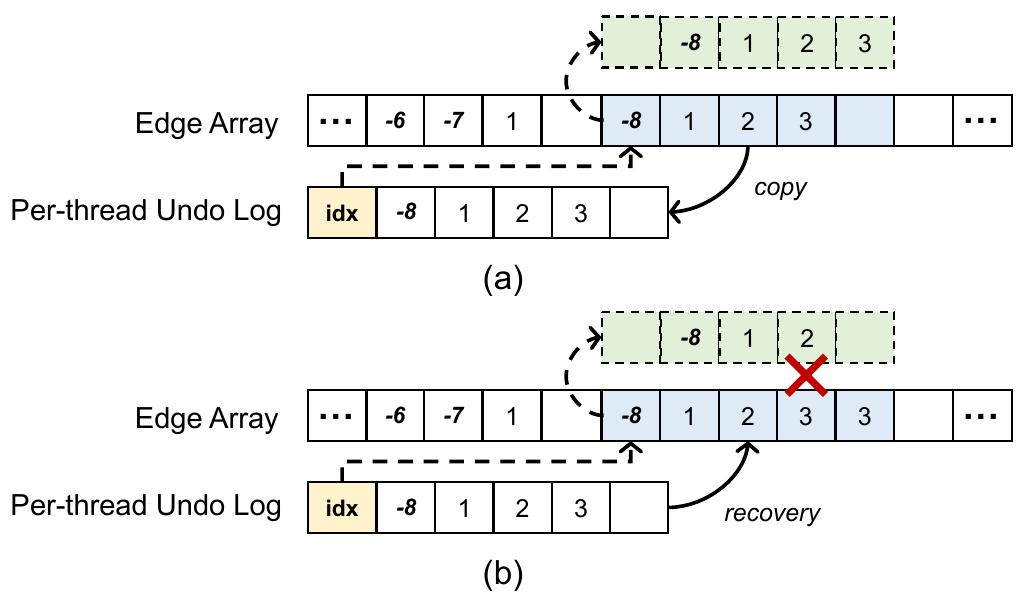}
	\caption{\small Crash consistent PMA \textit{rebalancing} in DGAP. In (a), the blue area shows the intended data movement region; the green dashed boxes show the expected state after the data movement. In (b), a crash case is shown after moving data $3$.}
	\label{fig:rebal}
\end{figure}




For every \textit{Writer Thread} in DGAP, a \textit{per-thread undo log} is allocated on PMs to support the execution of triggered \textit{rebalancing}.
Fig.~\ref{fig:rebal} illustrates the rebalancing process in detail. Once DGAP determines a valid rebalancing range based on density thresholds, it recalculates the location of each vertex within these sections, assuming the gaps will be redistributed evenly. For instance, in Fig.~\ref{fig:rebal}(a), the new location of vertex $v_8$ and its edges is represented by dashed boxes above the edge array. During rebalancing, all vertices and their edges must be moved to their new locations. To prevent permanent data loss in the event of a crash, this relocation process must be safeguarded using a transaction mechanism.

To perform data movements in a crash-consistent manner, DGAP first backs up the data that may be overwritten during data movements in the \textit{undo log}. It then calls \texttt{CLWB} and \texttt{SFENCE} to ensure that the data is persisted before proceeding with the actual data movement on the edge array. If a crash occurs before the backup of data on the undo log is completed, the data on the original edge array remains unaffected, as no data movement has occurred yet.
After the backup, DGAP initiates the process of moving and overwriting data element by element. DGAP iteratively performs these steps until the entire rebalancing range is moved. In each step, it moves a maximum of \texttt{ULOG\_SZ}=2K bytes of data.


Figure \ref{fig:rebal}(b) illustrates a crash scenario during rebalancing. In this instance, DGAP has already backed up the moving data in the \textit{undo log} and is beginning to shift all edges of $v_8$ one element to the right. Suppose that after the edge ($8,3$) has been moved, a crash occurs, resulting in an inconsistent \textit{edge array} due to the presence of two edges ($8, 3$). However, a consistent backup of this region is available in the persistent \textit{undo log}. 
Upon restart, DGAP recognizes the crash by checking its \texttt{NORMAL\_SHUTDOWN} flag. It then iterates through all \textit{per-thread undo logs} and utilizes the backup data to overwrite the inconsistent regions. The \textit{idx} index, stored at the beginning of the \textit{per-thread undo log}, is used to determine which part of the edge array should be overwritten for recovery. After restoring the data, DGAP proceeds to reissue the rebalancing operation to complete the interrupted process.

\subsubsection{Shutdown and Crash Recovery}
\label{sec:operation:reboot}
DGAP can initiate a graceful shutdown by calling \texttt{g.shutdown()}. During a normal shutdown, DGAP first waits for all ongoing graph analytic tasks to complete. Subsequently, it persists all DRAM components to persistent memory (PM), including the \textit{vertex array} and PMA-related metadata. While this backup process may require a few seconds, it ensures a quicker subsequent startup. Detailed normal shutdown times are measured and presented in the evaluation section. Before shutting down, DGAP resets the \texttt{NORMAL\_SHUTDOWN} flag to indicate a graceful shutdown.
	
After rebooting, DGAP first checks the \texttt{NORMAL\_SHUTDOWN} flag to understand whether the previous shutdown is normal or due to a crash. If the flag indicates a normal shutdown, DGAP simply loads the \textit{vertex array} and PMA-related metadata to DRAM and starts operating. If this is a reboot after a crash, DGAP initiates a data recovery process. Initially, DGAP scans the \textit{edge array} to reconstruct the \textit{vertex array} and build PMA metadata, such as the density tree. 
Following that, DGAP scrutinizes all \textit{per-thread undo logs} and recovers the inconsistencies resulting from crashed rebalancing operations. It then continues to finish the ongoing rebalancing from the inconsistent region.
Next, DGAP checks the \textit{per-section edge log} to retrieve the metadata for these vertices and update the \textit{vertex array}. 
After all these steps, DGAP can start to operate normally. 
In the evaluation section, we present the time durations associated with both standard and crash reboots.


\begin{table*}[h]
\caption{\small A list of graph kernels and inputs and outputs used in our evaluations.}
\small
\resizebox{\textwidth}{!}{%
    \centering
		\begin{tabular}{llcll}
			\toprule
			Graph kernel & Kernel Type & Input & Output & Notes\\
			\midrule
			PageRank (PR) & Link Analysis & - & $|V|$-sized array of ranks & Fixed number (20) of iterations\\
			Breadth-First Search (BFS) & Graph Traversal & Source vertex & $|V|$-sized array of parent IDs & Direction-Optimizing approach~\cite{beamer2012direction}\\
			Betweenness Centrality (BC) & Shortest Path & Source vertex & $|V|$-sized array of centrality scores & Brandes approx. algorithm~\cite{brandes2001faster, madduri2009faster}\\
			Connected Components (CC) & Connectivity & - & $|V|$-sized array of component labels & Shiloach-Vishkin~\cite{shiloach1980log, bader2005architectural}\\
			\bottomrule
		\end{tabular}
  }
		\label{tab:gkernels}
  \vspace{-1em}
\end{table*}

\begin{table}[h!]
    \caption{\small Graph inputs and their key properties.}
    \small
    \resizebox{\columnwidth}{!}{%
    \centering
		\begin{tabular}{llrrr}
			\toprule
			Datasets & Domain & $|V|$ & $|E|$ & $|E| / |V|$\\ 
			\midrule
			Orkut & social & 3,072,626 & 234,370,166 & 76\\ 
			LiveJournal & social & 4,847,570 & 85,702,474 & 18\\ 
			CitPatents & citation & 6,009,554 & 33,037,894 & 6\\ 
			Twitter & social & 61,578,414 & 2,405,026,390 & 39\\ 
            Friendster & social & 124,836,179 & 3,612,134,270 & 29\\
            Protein & biology & 8,745,543 & 1,309,240,502 & 149\\
			\bottomrule
		\end{tabular}
  }
		\label{tab:graphs}
  \vspace{-1em}
\end{table}

\subsubsection{Concurrency Control}
\label{sec:operation:concur}

DGAP supports multi-thread graph updates (multiple \textit{Writer Threads}) and graph analysis (multiple \textit{Analysis Tasks}) on PMs. To optimize performance, DGAP implements an optimistic read/write lock to enable multiple readers and writers to run concurrently, as long as they do not write to the same section.
For each PMA section, 
DGAP maintains a lock and its linked condition variable, resulting in $|log(v)|$ locks. 
When inserting an edge ($v_{src},v_{dst}$), DGAP first needs to acquire the lock for the respective section of ${v_{src}}$ so that no other threads can insert into the same section. This also prevents concurrent readers. After the insertion, DGAP checks whether the density of $Section_{v_{src}}$ has reached the rebalancing threshold. If rebalancing is needed, the writer thread first sets the condition variable of $Section_{v_{src}}$ to block other writes or rebalancing operations in this section. It then attempts to acquire all the locks of the sections affected by the rebalancing, sequentially. To prevent deadlocks, DGAP follows a strict order (from low to high section IDs) when acquiring locks. After obtaining all the locks, DGAP executes the rebalancing as previously described. Finally, DGAP resets the condition variable and notifies all waiting writes or rebalancing operations to start.
Note that DGAP stores all the locks in DRAM instead of PMs to increase performance. If a crash occurs, all the locks are lost. The pending rebalancing operation will be recovered by checking the \textit{per-thread undo log}. The pending edge writes will be ignored, as they have not yet been returned successfully to users.

\section{Evaluation}
\label{sec:eval}

We developed DGAP using the PMDK library~\cite{pmdk}. Its core data structure consists of approximately 2,000 lines of C++ code. The code is publicly available on Github\footnote{https://github.com/DIR-LAB/DGAP}. In this section, we compare DGAP with other graph analysis frameworks on real-world graphs with synthetic graph insertion patterns. The results reported are the averages of five runs.

\subsection{Evaluation Setup}
\label{sec:eval:setup}
\textbf{Evaluation Platform.} 
We conducted all evaluations on a Dell R740 rack server equipped with a 2nd generation Intel Xeon Scalable Processor (Gold 6254 @ 3.10 GHz) featuring 18 physical cores. The server also included 6 DRAM DIMMs with 32 GB each (for a total of 192 GB) and 6 Optane DC DIMMs with 128 GB each (for a total of 768 GB). We configured Optane DC in \textit{App Direct} mode. The system ran Ubuntu 20.04 and used the Linux kernel version 4.15.0. Our implementation is based on PMDK 1.12.

\textbf{Graph Algorithms.}
To ensure a fair comparison among various graph analysis frameworks, we used the same implementations of four graph algorithms from the GAP Benchmark Suite (GAPBS)~\cite{gapbs}. These algorithms are PageRank (PR), Breadth-First Search (BFS), Betweenness Centrality (BC), and Connected Components (CC), detailed in Table~\ref{tab:gkernels}. GAPBS also offers an optimized Compressed Sparse Row (CSR) implementation, which we modified for persistent memory to serve as one of our evaluation baselines.

\textbf{Graph Datasets.} 
We used several real-world graphs from SNAP datasets~\cite{snapnets} in our evaluations. Table~\ref{tab:graphs} lists these graphs and their key properties. We generate the insertion order by randomly shuffling all the edges for these datasets.
Note that, in all the experiments, we will insert the first \textit{10}\% of the graph and then start to benchmark the insertion performance for the purpose of warming up the system, similar to the warm-up stage in YCSB~\cite{cooper2010benchmarking}. 

\textbf{Compared Systems.}
To showcase the performance of DGAP, we compare it with multiple data structures and state-of-the-art dynamic graph frameworks. 

First, we ported two foundational graph data structures to persistent memory to serve as baselines. The \textbf{Compressed Sparse Row (CSR)} on persistent memory is based on GAPBS. CSR serves as a baseline for graph analysis evaluations since 1) it can not be updated and 2) it offers the optimal graph analysis performance due to its compact memory layout.
We also implemented \textbf{Blocked Adjacency-List (BAL)} on persistent memory as another extreme baseline. BAL is known to have poor graph analysis performance due to pointer chasing and great edge insertions performance due to efficient appending to a block. We use BAL as a baseline to understand the insertion performance of DGAP.

We further compared DGAP with three state-of-the-art dynamic graph frameworks designed to support graph updates and analysis. 
\textbf{LLAMA} uses a multi-versioned CSR structure to enable fast graph analysis and graph mutations~\cite{macko2015llama}. The graph updates are conducted in batches and organized as multiple immutable snapshots in LLAMA. 
To avoid creating too many snapshots, in our evaluation, we only created a snapshot after inserting $1\%$ of the graph, which ranges from 330K edges to 36M edges, depending on the chosen graph dataset. In total, we created 90 snapshots for each graph (the first 10\% warm-up is a single snapshot). Because graph analysis in LLAMA can not read the latest graph unless the snapshot is created, these large snapshots mean its graph analysis tasks may miss as many as 36 million edges, which might not be acceptable in some applications. 
We ported {{LLAMA}} to persistent memory by changing the location of its snapshot files to PMs space, which shows a naive way of moving existing graph data structure to persistent memory.

\textbf{GraphOne} is an in-memory graph analysis framework with an extra durability guarantee using external non-volatile devices~\cite{kumar2019graphone}. 
New data is first stored in an in-DRAM edge list in an append-only manner. Background threads incrementally move this data to non-volatile memory for persistence. 
To port GraphOne to persistent memory, we changed the location of \textit{durable phase} to the PM space and required it to flush DRAM data after each $2^{16}$ insertions to reduce the chances of losing data. We do not limit the DRAM usage of GraphOne during graph analysis. Hence for some graphs, the graph data may be completely cached in DRAM. Due to these settings, we name this baseline as \textbf{GraphOne-FD}, indicating GraphOne Flushing-DRAM, in the rest of the paper.

\begin{figure}[t]
    \centering
    \begin{subfigure}[b]{1.0\linewidth}
    	\centering
    	\includegraphics[page=1,width=1.0\linewidth]{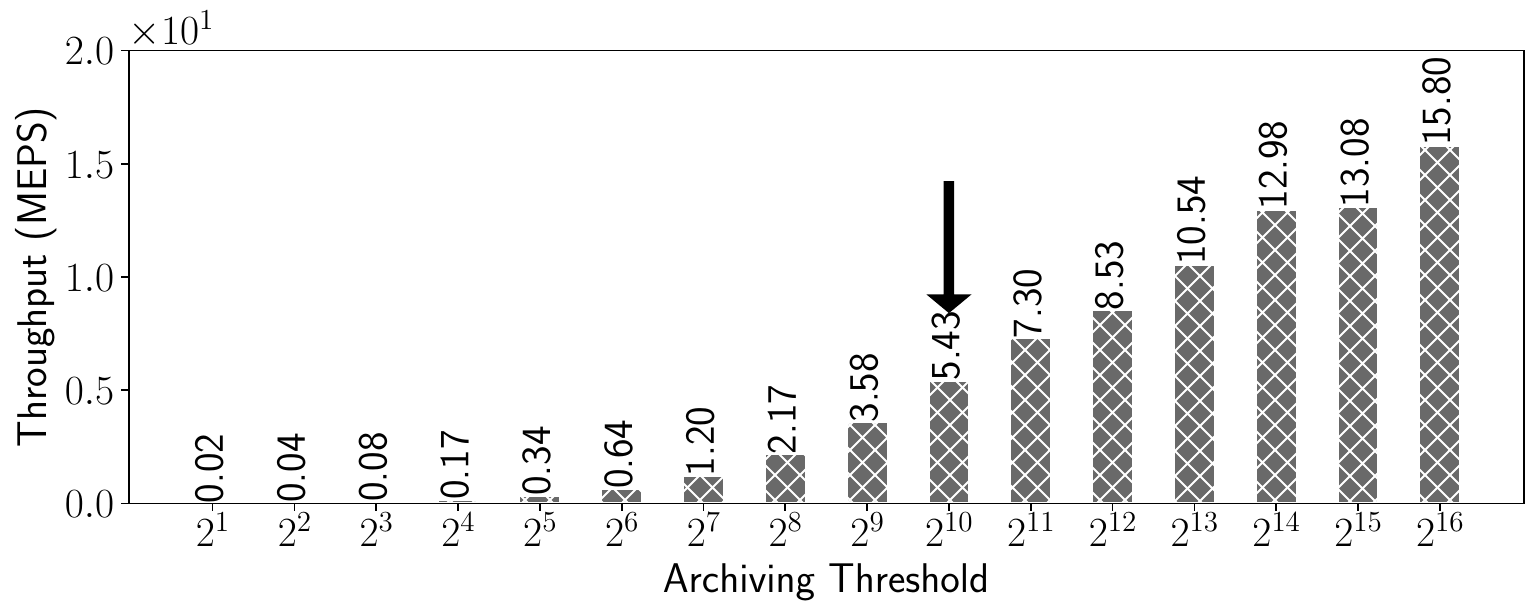}
    	\label{fig:sub-first}
    \end{subfigure}
    \caption{\small Graph insertion throughput of XPGraph (in MEPS) for varying \textit{Archiving Threshold}. \textit{Higer value is better.}}
    \label{eval:insert_xpgraph}
\end{figure}

\textbf{XPGraph} is state-of-the-art PM-based dynamic graph system~\cite{wang2022xpgraph}. It is based on GraphOne but extends it with new designs for persistent memory. Specifically, XPGraph stores both the edge list and adjacency list in persistent memory to guarantee data persistence and leverages the DRAM as a cache to batch data into the adjacency list. Similar to GraphOne, XPGraph also transfers data to DRAM for graph analysis. In our evaluations, we used the default parameter settings of XPGraph for comparisons. \arxiv{One exception is that we used a lower \textit{archiving threshold} to record the graph insert performance of XPGraph. Here, \textit{archiving threshold} indicates the batch size of edges to move from the edge list to the adjacency list. In theory, XPGraph is supposed to perform both operations (adding new edge to edge list and moving the edges to adjacency list) concurrently, which would incur higher overhead due to cache line inference. But, in their current prototype implementation~\cite{xpgraph-repo}, XPGraph performed these operation separately and avoided the overhead. The insert performance of XPGraph thus solely depends on the \textit{archiving threshold} as showing in Fig.~\ref{eval:insert_xpgraph}. For a fair comparison, we picked threshold $2^{10}$ in our evaluations, which still means XPGraph may delay the real-time analysis by $2^{10}$ edges.}

  

\subsection{Graph Insertions Performance}

\begin{figure}[t]
    \centering
    \begin{subfigure}[b]{1.0\linewidth}
    	\centering
    	\includegraphics[page=2,width=0.9\linewidth]{exps/SC23_0__single_thread_insert_throughput.pdf}
    	\label{fig:sub-first}
    \end{subfigure}
    \begin{subfigure}[b]{1.0\linewidth}
    	\centering
    	\includegraphics[page=1,width=\linewidth]{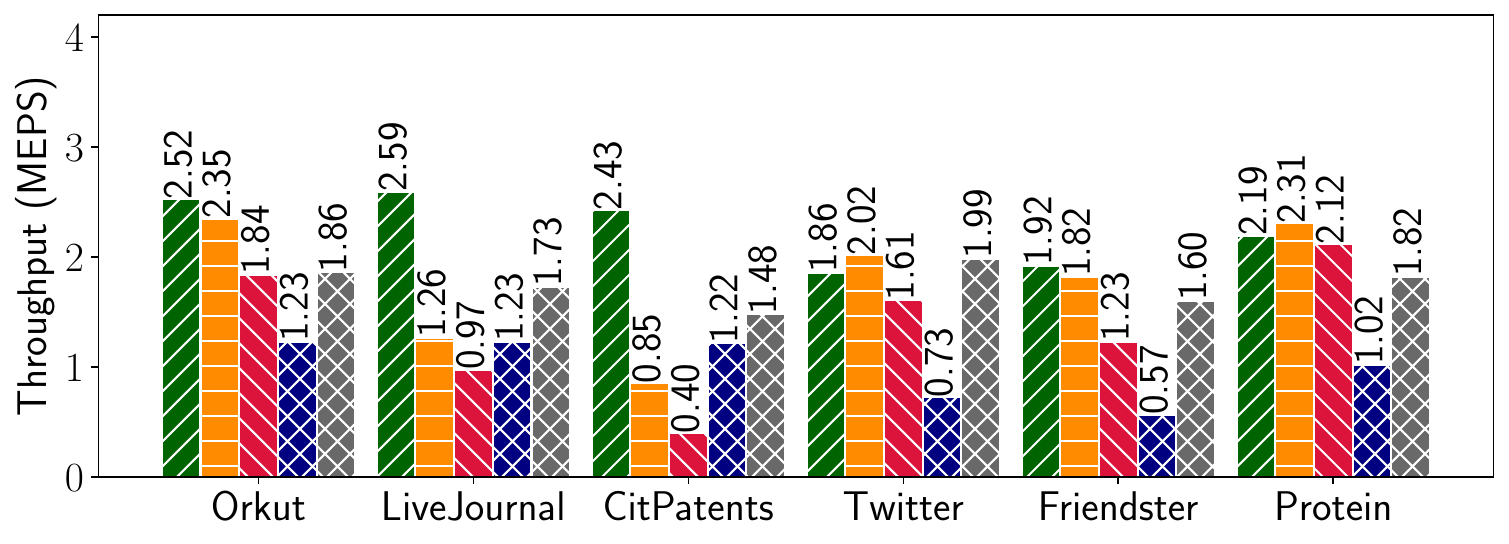}
    	\label{fig:sub-second}
    \end{subfigure}
    \caption{\small Dynamic graph insertion throughput in million edges per second (MEPS). \textit{Higer value is better.}}
    \label{eval:insert_single}
\end{figure}

We first compared the graph updates, particularly the edge insertion performance of DGAP, with other systems.
Fig.~\ref{eval:insert_single} shows the graph insertion throughput in MEPS (Million Edges Per Second) using a single writer thread. The scalability results are reported later. From these results, we can observe that DGAP achieves almost the best performance across all datasets among all the frameworks. 
It delivers $1.03\times-2.82\times$ better performance than BAL, which is considered extremely efficient in graph insertions as edges are simply appended to the end of each block. However, the inefficient usage of persistent memory (e.g., journaling and transaction for crash consistency) makes it slower in many cases. 
DGAP also outperforms LLAMA, GraphOne, and XPGraph on persistent memory by up to $6\times$, $2.5\times$, and $2.3\times$, respectively. It is obvious to us that the costs of asynchronous batch data structure conversions and movements between DRAM and PMs in LLAMA, GraphOne, and XPGraph impact the performance significantly. It is worth noting that, from the results, XPGraph performs better than GraphOne, but not as significant as the original paper reports~\cite{wang2022xpgraph}. This is because our GraphOne-FD has a large batch write size in DRAM, which offers better performance but is impractical as this data may be lost. Still, the better performance of DGAP clearly showcases the efficiency of mutable CSR data structure on persistent memory.

\begin{table*}[h!]
    \caption{Graph insertion throughput (MEPS) using the different number of writer threads. \textit{Larger throughput is better.}}
    \resizebox{\textwidth}{!}{%
	\centering
	\begin{tabular}{l arrrrarrrrarrrr}
        \toprule
        & \multicolumn{5}{c}{$T_1$} & \multicolumn{5}{c}{$T_8$} & \multicolumn{5}{c}{$T_{16}$}\\
        \cmidrule(lr){2-6}
        \cmidrule(lr){7-11}
        \cmidrule(lr){12-16}
        Graph     & DGAP & BAL & LLAMA & GO-FD & XPGrp. & DGAP & BAL & LLAMA & GO-FD & XPGrp. & DGAP & BAL & LLAMA & GO-FD & XPGrp.\\
        \midrule

        
        Orkut
        & \B\num{2.524614111} & \num{2.34996497} & \num{1.840474923} & \num{1.23074542} & \num{1.863023076}
        & \B\num{6.486952403} & \num{5.97307019} & \num{2.331912553} & \num{2.53638064} & \num{4.949091481}
        & \B\num{7.37293903} & \num{5.261014687} & \num{2.395361633} & \num{2.862305266} & \num{5.436309287}\\
        
        LiveJournal
        & \B\num{2.593619269} & \num{1.264353402} & \num{0.9730563027} & \num{1.227506247} & \num{1.731583494}
        & \B\num{6.265831328} & \num{4.786963736} & \num{1.073308291} & \num{2.63362727} & \num{4.917770254}
        & \B\num{7.953504022} & \num{5.920610006} & \num{1.086951143} & \num{2.936533122} & \num{5.659880334}\\
        
        CitPatents
        & \B\num{2.426446905} & \num{0.8531981547} & \num{0.3976955434} & \num{1.215493667} & \num{1.481247573}
        & \B\num{6.817161146} & \num{3.451580363} & \num{0.41229797} & \num{2.622529205} & \num{5.045316673}
        & \B\num{7.233994594} & \num{4.680063525} & \num{0.4158325292} & \num{2.807454827} & \num{5.753064608}\\
        
        Twitter
        & \num{1.858658294} & \B\num{2.017550258} & \num{1.605122506} & \num{0.7251165802} & \num{1.986590775}
        & \num{5.3489936} & \B\num{5.507841176} & \num{2.129158429} & \num{1.992767149} & \num{4.881072383}
        & \B\num{6.821753892} & \num{5.987000648} & \num{2.171520389} & \num{2.433096893} & \num{5.326770861}\\
        
        Friendster
        & \B\num{1.920984709} & \num{1.818648647} & \num{1.23099628} & \num{0.5675252507} & \num{1.60123691}
        & \num{4.291106517} & \B\num{5.626404144} & \num{1.520849545} & \num{2.404349414} & \num{4.40805217}
        & \B\num{6.025289488} & \num{5.820030886} & \num{1.533593566} & \num{3.345911263} & \num{4.997107637}\\
        
        Protein
        & \num{2.193612403} & \B\num{2.305273709} & \num{2.117883855} & \num{1.01926962} & \num{1.819069211}
        & \B\num{7.429157925} & \num{5.822706911} & \num{3.086205775} & \num{3.214630692} & \num{5.082988132}
        & \B\num{8.298061613} & \num{6.225190198} & \num{3.183228112} & \num{4.0782497} & \num{5.759612263}\\
        
        \bottomrule
    \end{tabular}
    }
    
    \label{tab:scalable_write}
\end{table*}

\subsubsection{Graph Insertions Scalability}
We further evaluated the graph insertions scalability by increasing the number of concurrent writer threads from 1 to 16. Table~\ref{tab:scalable_write} shows the MEPS throughput of 1, 8, and 16 threads.
We can see DGAP scales with more threads. It delivers up to $4.3\times$ throughput in $16$ threads compared with single thread case. 
The concurrency model and write optimizations implemented in DGAP help deliver such a scalable graph insertion performance ($6$ to $8$ million edges/sec), which might be needed in many real-time big data applications. 




Across various systems, DGAP consistently ranks as either the best or very close to the best in all scalability cases. BAL occasionally delivers superior performance, primarily due to our implementation of BAL utilizing finer-grain locks for concurrent insertions. Specifically, while DGAP locks writers by edge section, BAL employs vertex-based locking. Consequently, as the number of threads increases, its performance scales more effectively. However, this may not be a realistic representation, as an excessive number of locks are needed.
The scalability results of XPGraph are also noteworthy, as it surpasses DGAP in the 16-thread case for three graphs. In fact, these three graphs are all relatively small. We attribute the exceptional performance to XPGraph's design. Specifically, XPGraph includes a circular edge log for temporarily storing new insertions. By default, the circular edge log has a capacity of 8GB, which can entirely accommodate the three smaller graphs: Orkut, LiveJournal, and CitPatents. In this context, archiving is not activated for these graphs, resulting in XPGraph exhibiting exceptional performance. For larger graphs with over a billion edges, DGAP demonstrates $12-21\%$ better performance, as XPGraph is compelled to flush the DRAM caches back to the persistent edge list more frequently.

\subsection{Graph Analysis Performance}

\begin{figure}[t]
    \centering
    \begin{subfigure}[b]{1.0\linewidth}
    	\centering
    \includegraphics[page=2,width=\linewidth]{exps/SC23_0__single_thread_nomalized_pr.pdf}
    	\label{fig:sub-first}
    \end{subfigure}
    \begin{subfigure}[b]{1.0\linewidth}
    	\centering
    	\includegraphics[page=1,width=\linewidth]{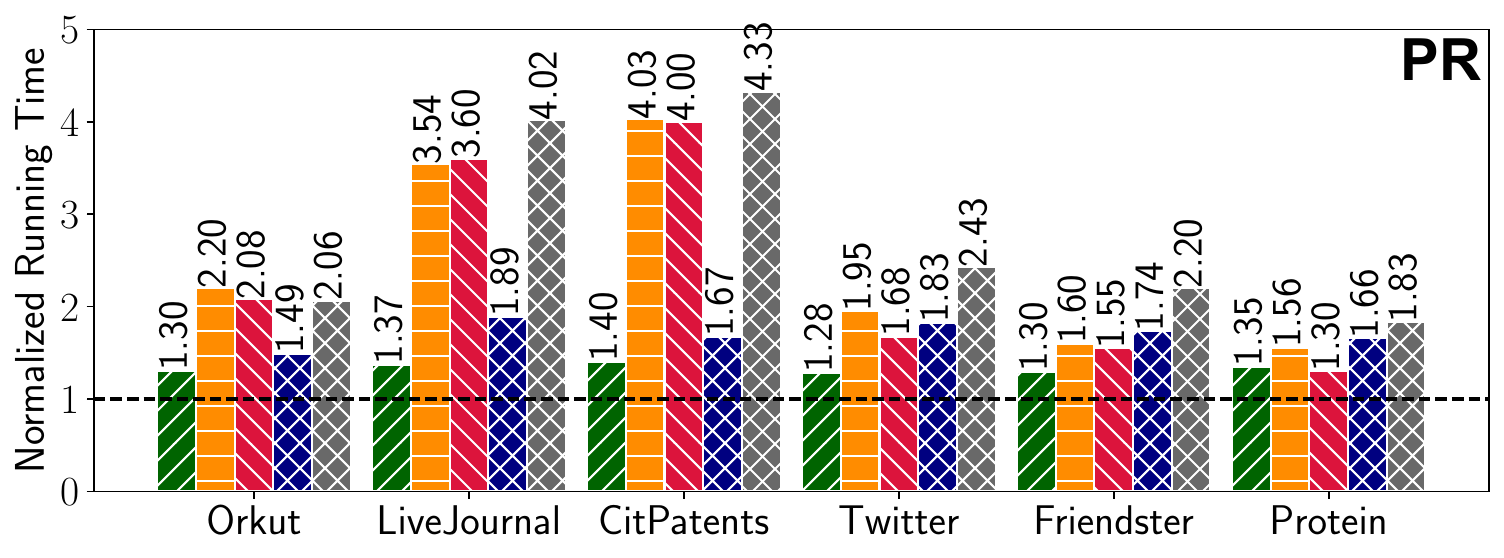}
    	\label{fig:sub-second}
    \end{subfigure}
    \begin{subfigure}[b]{1.0\linewidth}
    	\centering
    	\includegraphics[page=1,width=\linewidth]{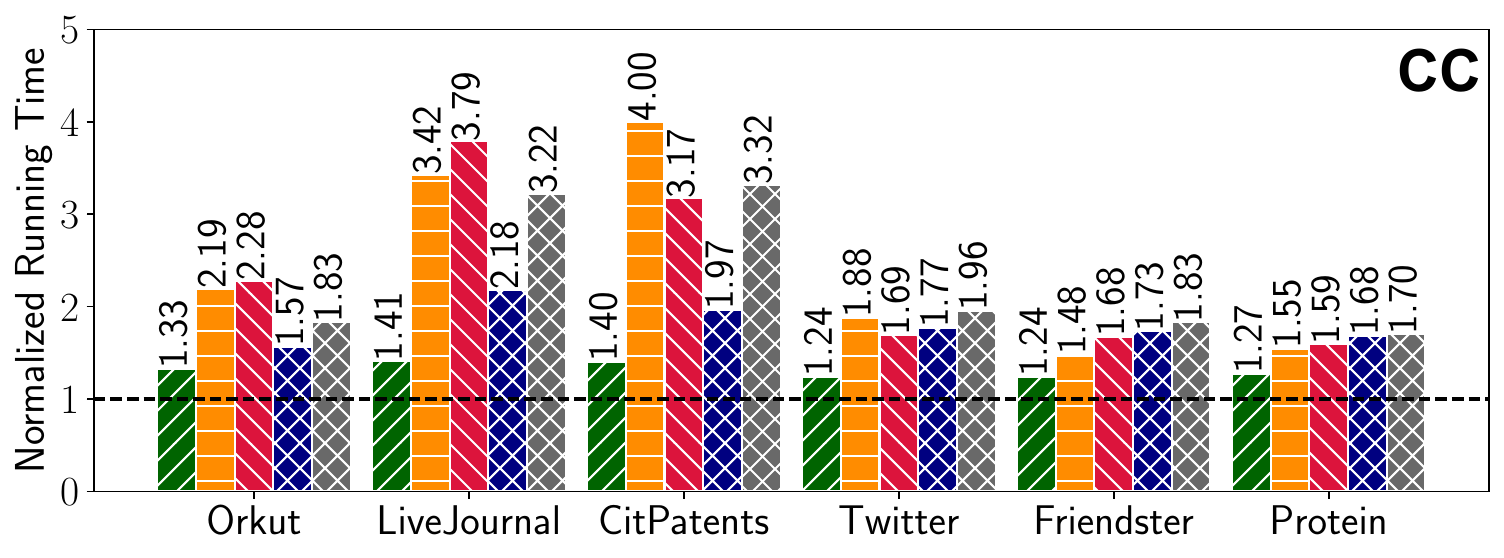}
    	\label{fig:sub-second}
    \end{subfigure}
    \caption{\small Time to run PageRank (PR) and Connected Components (CC), normalized to CSR on PMM. \textit{Smaller is better.}}
    \label{eval:pr_single}
\end{figure}

Graph analysis performance is key to our graph frameworks. In this section, we show the performance of running four classic graph algorithms (listed in Table~\ref{tab:gkernels}) on different graphs. Among these four algorithms, PageRank (PR) and Connected Components (CC) access all vertices in each iteration, while Breadth-First Search (BFS) and Betweenness Centrality (BC) access parts of the graphs each time based on the calculation. They show different access patterns which may impact the performance of the frameworks, as shown below. 




\textit{1) PageRank (PR) and Connected Components (CC).}
Fig.~\ref{eval:pr_single} illustrates the relative speed of PageRank compared to CSR using a single thread. Compared with CSR, which is best for graph analysis, DGAP introduces only $37\%$ overhead on average and achieves up to $2.9\times$, $2.9\times$, $1.4\times$, and $3.1\times$ better performance compared to BAL, LLAMA, GraphOne, and XPGraph respectively.
It is particularly interesting to observe that DGAP outperforms GraphOne-FD in most datasets, even it is actually running on DRAM-cached data. We believe this is because GraphOne uses adjacency list as its in-memory data structure, which is less efficient for graph analysis tasks that apply to all vertices and edges of the graph. While, since DGAP is a mutable CSR, it shows much better cache locality in running these algorithms such as PageRank.
We can observe the same behaviors when running a similar algorithm CC, which iterates all vertices/edges in each iteration. Specifically, Fig.~\ref{eval:pr_single} illustrates the relative speed of CC compared to CSR on all the systems. Again, DGAP shows up to $2.9\times$, $2.7\times$, $1.6\times$, and $2.4\times$ better performance than BAL, LLAMA, GraphOne, and XPGraph respectively.

\begin{figure}[t]
    \centering
    \begin{subfigure}[b]{1.0\linewidth}
    	\centering
    	\includegraphics[page=2,width=\linewidth]{exps/SC23_0__single_thread_nomalized_bfs.pdf}
    	\label{fig:sub-first}
    \end{subfigure}
    \begin{subfigure}[b]{1.0\linewidth}
    	\centering
    	\includegraphics[page=1,width=\linewidth]{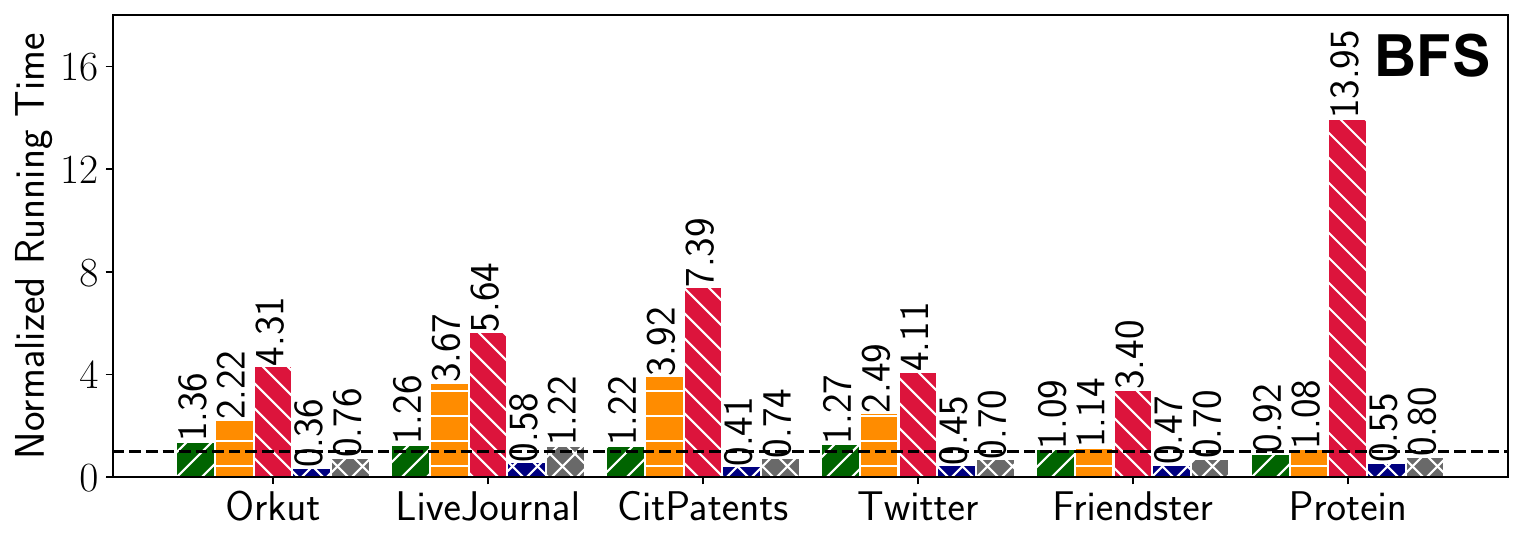}
    	\label{fig:sub-second}
    \end{subfigure}
    \begin{subfigure}[b]{1.0\linewidth}
    	\centering
    	\includegraphics[page=1,width=\linewidth]{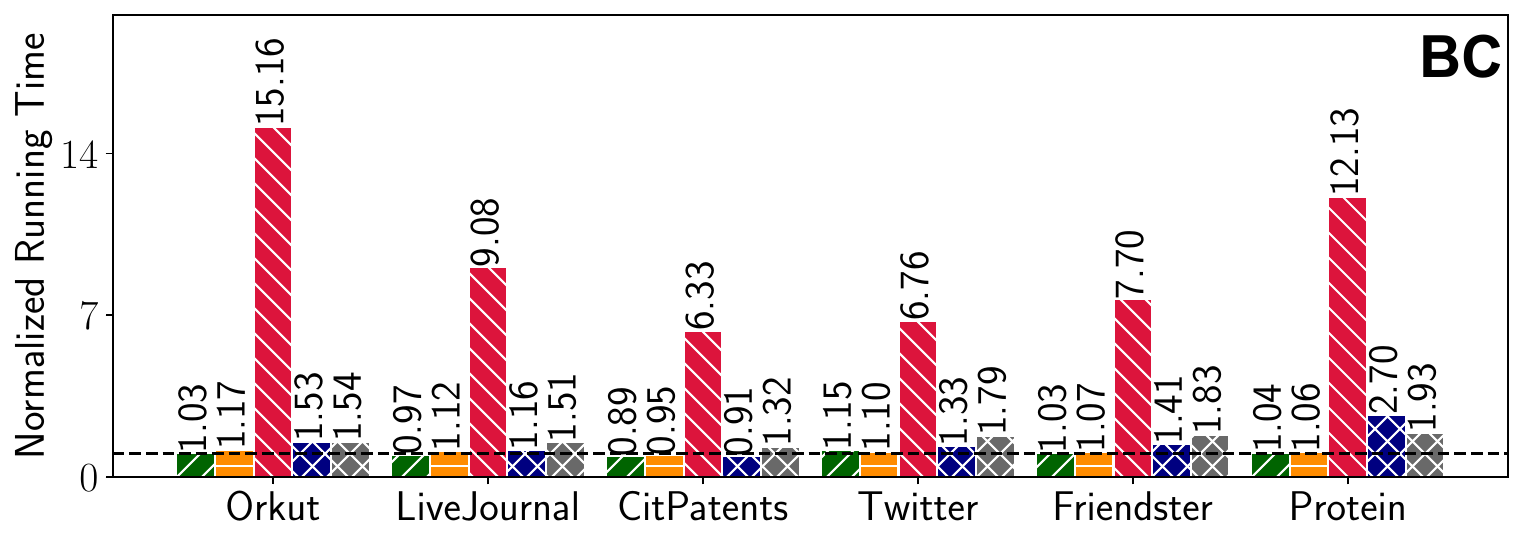}
    	\label{fig:sub-second}
    \end{subfigure}
    \caption{\small Time to run Breadth-First Search (BFS) and Betweenness Centrality (BC), normalized to CSR on PMM. \textit{Smaller is better.}}
    \label{eval:bfs_single}
    \vspace{-1em}
\end{figure}

\begin{table*}[h!]
    \caption{\small The execution time (in seconds) of four algorithms on all systems. $T_1$ denotes the time of one thread and $T_{16}$ denotes that of 16 threads.}
    \resizebox{\textwidth}{!}{%
	\centering
	\begin{tabular}{l aabbrrrrrrrraabbrrrrrrrr}
        \toprule & \multicolumn{12}{c}{PageRank} & \multicolumn{12}{c}{BFS} \\
        \cmidrule(lr){2-13}
        \cmidrule(lr){14-25}
        & \multicolumn{2}{c}{CSR} & \multicolumn{2}{c}{DGAP} & \multicolumn{2}{c}{BAL} & \multicolumn{2}{c}{LLAMA} & \multicolumn{2}{c}{GraphOne} & \multicolumn{2}{c}{XPGraph} & \multicolumn{2}{c}{CSR} & \multicolumn{2}{c}{DGAP} & \multicolumn{2}{c}{BAL} & \multicolumn{2}{c}{LLAMA} & \multicolumn{2}{c}{GraphOne} & \multicolumn{2}{c}{XPGraph}\\
        Graph     & $T_1$ & $T_{16}$ & $T_1$ & $T_{16}$ & $T_1$ & $T_{16}$ & $T_1$ & $T_{16}$ & $T_1$ & $T_{16}$ & $T_1$ & $T_{16}$ & $T_1$ & $T_{16}$ & $T_1$ & $T_{16}$ & $T_1$ & $T_{16}$ & $T_1$ & $T_{16}$ & $T_1$ & $T_{16}$ & $T_1$ & $T_{16}$\\
        \midrule
        
        Orkut          & \num{24.17801}	& \num{1.66891}	& \num{31.55084}	& \num{2.20741}	& \num{53.21021}	& \num{3.56587}	& \num{50.241}	& \num{9.514}	& \num{36.0078}	& \num{2.62825} & \num{49.86686} & \num{3.72478}	& \num{0.33473}	& \num{0.03006}	& \num{0.45669}	& \num{0.03858}	& \num{0.74455}	& \num{0.05673}	& \num{1.444}	& \num{0.331}	& \num{0.119332}	& \num{0.012429} & \num{0.2528406} & \num{0.02819198}\\

        
        LiveJournal    & \num{9.0714}	& \num{0.70642}	& \num{12.46027}	& \num{0.93678}	& \num{32.11773}	& \num{2.30393}	& \num{32.685}	& \num{5.124}	& \num{17.1449}	& \num{1.23757} & \num{36.45106} & \num{3.035008}	& \num{0.34283}	& \num{0.03262}	& \num{0.43279}	& \num{0.03734}	& \num{1.25658}	& \num{0.10219}	& \num{1.934}	& \num{0.504}	& \num{0.197539}	& \num{0.025619} & \num{0.419764} & \num{0.04820182}\\
        
        CitPatents     & \num{5.82708}	& \num{0.4867}	& \num{8.17053}	& \num{0.63161}	& \num{23.47029}	& \num{1.72618}	& \num{23.304}	& \num{2.826}	& \num{9.75156}	& \num{0.703277} & \num{25.21416} & \num{2.377254}	& \num{0.4688}	& \num{0.04374}	& \num{0.57417}	& \num{0.04716}	& \num{1.83742}	& \num{0.13827}	& \num{3.464}	& \num{0.675}	& \num{0.194491}	& \num{0.025049} & \num{0.3470958} & \num{0.05607478}\\

        Twitter     & \num{425.1114}	& \num{31.58598}	& \num{545.92395}	& \num{39.3025}	& \num{828.06803}	& \num{56.6687}	& \num{712.729}	& \num{99.829}	& \num{775.83}	& \num{45.0997} & \num{1032.062} & \num{77.988}	& \num{7.91179}	& \num{0.70703}	& \num{10.08591}	& \num{0.73936}	& \num{19.71679}	& \num{1.46779}	& \num{32.497}	& \num{6.649}	& \num{3.57858}	& \num{0.334013} & \num{5.54679} & \num{0.713467}\\

        Friendster	& \num{873.37641}	& \num{65.40583}	& \num{1131.83671}	& \num{80.8384}	& \num{1394.0507}	& \num{97.70416}	& \num{1353.574}	& \num{186.814}	& \num{1515.38}	& \num{85.7671} & \num{1922.262} & \num{142.493}	& \num{14.7672}	& \num{1.12491}	& \num{16.09514}	& \num{1.18811}	& \num{16.78743}	& \num{1.41422}	& \num{50.231}	& \num{13.54}	& \num{6.92292}	& \num{0.501637}  & \num{10.4098} & \num{1.07127}\\

        Protein	& \num{203.48426}	& \num{13.21742}	& \num{274.91033}	& \num{16.85049}	& \num{316.64591}	& \num{20.42522}	& \num{264.232}	& \num{34.586}	& \num{336.888}	& \num{20.6053} & \num{372.1148} & \num{27.9637}	& \num{0.89642}	& \num{0.08315}	& \num{0.82143}	& \num{0.08003}	& \num{0.96726}	& \num{0.09896}	& \num{12.505}	& \num{1.266}	& \num{0.495898}	& \num{0.0445359}   & \num{0.7153644} & \num{0.08693724}\\
        
        \bottomrule
    \end{tabular}
    }

\resizebox{\textwidth}{!}{%
	\centering
    \begin{tabular}{l aabbrrrrrrrraabbrrrrrrrr}
        & \multicolumn{12}{c}{BC} & \multicolumn{12}{c}{CC} \\
        \cmidrule(lr){2-13}
        \cmidrule(lr){14-25}
        & \multicolumn{2}{c}{CSR} & \multicolumn{2}{c}{DGAP} & \multicolumn{2}{c}{BAL} & \multicolumn{2}{c}{LLAMA} & \multicolumn{2}{c}{GraphOne} & \multicolumn{2}{c}{XPGraph} & \multicolumn{2}{c}{CSR} & \multicolumn{2}{c}{DGAP} & \multicolumn{2}{c}{BAL} & \multicolumn{2}{c}{LLAMA} & \multicolumn{2}{c}{GraphOne} & \multicolumn{2}{c}{XPGraph}\\
        Graph     & $T_1$ & $T_{16}$ & $T_1$ & $T_{16}$ & $T_1$ & $T_{16}$ & $T_1$ & $T_{16}$ & $T_1$ & $T_{16}$ & $T_1$ & $T_{16}$ & $T_1$ & $T_{16}$ & $T_1$ & $T_{16}$ & $T_1$ & $T_{16}$ & $T_1$ & $T_{16}$ & $T_1$ & $T_{16}$ & $T_1$ & $T_{16}$\\
        \midrule
        
        Orkut           & \num{5.21541}	& \num{0.42425}	& \num{5.39668}	& \num{0.41732}	& \num{6.10334}	& \num{0.45767}	& \num{79.067}	& \num{5.714}	& \num{7.97608}	& \num{0.581536} & \num{8.008688} & \num{0.8059866}	& \num{2.60356}	& \num{0.4192}	& \num{3.45188}	& \num{0.7272}	& \num{5.71225}	& \num{0.88495}	& \num{5.94}	& \num{0.865}	& \num{4.08132}	& \num{0.751085} & \num{4.766} & \num{0.7106646}\\


        LiveJournal    & \num{4.37368}	& \num{0.33103}	& \num{4.23143}	& \num{0.31824}	& \num{4.90669}	& \num{0.36021}	& \num{39.719}	& \num{2.759}	& \num{5.05696}	& \num{0.358306} & \num{6.619362} & \num{0.609339}	& \num{0.9926}	& \num{0.42197}	& \num{1.39899}	& \num{0.79912}	& \num{3.39929}	& \num{0.8714}	& \num{3.764}	& \num{1.17}	& \num{2.16498}	& \num{0.749849} & \num{3.196488} & \num{1.026591}\\

        CitPatents     & \num{3.90332}	& \num{0.29427}	& \num{3.48582}	& \num{0.26086}	& \num{3.70786}	& \num{0.27233}	& \num{24.723}	& \num{1.698}	& \num{3.53534}	& \num{0.263158} & \num{5.153832} & \num{0.4650854}	& \num{1.66957}	& \num{0.48469}	& \num{2.34224}	& \num{0.48764}	& \num{6.68397}	& \num{1.42682}	& \num{5.296}	& \num{2.065}	& \num{3.2841}	& \num{0.809273} & \num{5.541868} & \num{1.677752}\\

        Twitter     & \num{106.09864}	& \num{7.83237}	& \num{122.38752}	& \num{7.86073}	& \num{117.08819}	& \num{8.41357}	& \num{717.385}	& \num{48.537}	& \num{141.165}	& \num{9.12626} & \num{190.2196} & \num{15.8141}	& \num{71.52673}	& \num{16.45203}	& \num{88.76109}	& \num{23.47609}	& \num{134.42366}	& \num{28.67994}	& \num{121.056}	& \num{25.199}	& \num{126.663}	& \num{24.8039} & \num{139.904} & \num{30.8892}\\

        Friendster	& \num{203.63407} & \num{14.69863} & \num{209.34011} & \num{14.46897} & \num{216.9206} & \num{15.14753} & \num{1568.58} & \num{105.367} & \num{287.513} & \num{17.5949} & \num{372.874} & \num{28.954} & \num{155.40058} & \num{23.72046} & \num{192.70532} & \num{36.40949} & \num{229.48163} & \num{33.45033} & \num{260.483} & \num{42.374} & \num{269.541} & \num{37.7942} & \num{284.5334} & \num{44.6524}\\

        Protein	& \num{2.01267} & \num{0.31297} & \num{2.09321} & \num{0.26585} & \num{2.14001} & \num{0.26546} & \num{24.418} & \num{1.861} & \num{5.43275} & \num{0.448226} & \num{3.879688} & \num{0.4729666} & \num{66.50267} & \num{4.51701} & \num{84.51543} & \num{6.74323} & \num{102.86298} & \num{6.7385} & \num{106.005} & \num{11.186} & \num{112.012} & \num{9.6705} & \num{113.2578} & \num{11.93266}\\
        
        \bottomrule
    \end{tabular}
    }
    
    \label{tab:actual_algo}
\end{table*}

\textit{2) Breadth-First Search (BFS) and Betweenness Centrality (BC).}
Fig.~\ref{eval:bfs_single} shows the relative speed of Breadth-First Search and Betweenness Centrality compared to CSR. For BFS, DGAP outperforms BAL and LLAMA by $2.30\times$ and $3.71\times$, respectively on average. However, DGAP performs $2.77\times$ and $1.81\times$ worse than GraphOne and XPGraph in this particular workload. This is expected since BFS is accessing edges of random vertices each time. The adjacency list in GraphOne and XPGraph performs very well for these tasks. CSR can not fully leverage its own spatial locality. In addition, since most BFS only reaches a small part of the graph, GraphOne and XPGraph can successfully cache the graph in DRAM. We observe similar trends for Betweenness Centrality (BC) as Fig.~\ref{eval:bfs_single} shows. Since BC is more computationally and memory intensive. It also covers larger parts of the graphs during computation, we can see that DGAP actually catches up and delivers similar performance compared with DRAM-based GraphOne and XPGraph.
Specifically, DGAP outperforms BAL, LLAMA, GraphOne, and XPGraph by up to $1.08\times$, $8.19\times$, $1.21\times$, and $1.85\times$ respectively. 

\subsubsection{Graph Analysis Scalability}
To examine the scalability of DGAP, we further ran the same graph algorithms using 1 to 16 threads and report the execution time (in seconds) in Table~\ref{tab:actual_algo}. Due to the space limits, we only report results of 1 thread and 16 threads for each case. 
From these results, we make server observations. 
First, DGAP scales well. It delivers up to $14.3\times$, $13.6\times$, $15.6\times$, and $4.7\times$ speedup using 16x threads running PageRank, BFS, BC, and CC algorithms respectively. It is interesting to see that DGAP does not scale well in CC. In fact, all the systems do not scale well in this algorithm. After checking the source code, we noticed the bottleneck actually comes from its inappropriate \textit{parallel for} scheduling keywords. If fixed, CC will deliver similar scalability for all frameworks. Since our goal is not to improve the algorithm implementation, we reported the results from the original GAPS implementation. Second, DGAP still delivers the best performance in most graph analysis algorithms. Similar to the single thread case, DGAP performs worse than GraphOne and XPGraph in the BFS case. As discussed earlier, this is mostly because GraphOne and XPGraph run BFS purely in DRAM and their adjacency list structure fits BFS well. 

\subsection{DGAP Components Evaluations}
\begin{table}[ht!]
    \small
    \caption{\small Insertion performance (in seconds) of different DGAPs.}
    \centering
		\begin{tabular}{lcccc}
			\toprule
			Datasets & DGAP & No EL & No EL\&UL & No EL\&UL\&DP\\
			\midrule
			Orkut & \num{83.55065} & 374.86 & 383.52 & 588.37\\
            LiveJournal    & \num{29.739225}  & \num{136.283} & \num{146.092} & \num{240.463}\\
            CitPatents & \num{12.254175} & \num{51.261} & \num{58.4692} & \num{107.388}\\
			\bottomrule
		\end{tabular}
		\label{tab:microbench_dgap_comp}
  \vspace{-1em}
\end{table}

\textbf{DGAP Components Evaluations.} In DGAP, we introduce three designs to maximize PMs. We further evaluated their contributions to the final performance. Specifically, we implemented and compared three different versions of DGAP by incrementally excluding its key components: (i) removing \textit{per-section Edge
Logs} as `No EL'; (ii) further removing \textit{per-thread Undo Log} as `No EL\&UL', replaced using PMDK transactions; and (iii) further removing Data Placement in DRAM as `No EL\&UL\&DP', meaning both vertex array and edge array are on PMs.
The graph insertion performance results are reported in Table~\ref{tab:microbench_dgap_comp}. We only report the results for small-size graphs, as we were not able to finish running all the tests on larger graphs in a reasonable time.



The results show that the \textit{per-section edge log} contributes the most in performance improvements. Without it, DGAP performs $4.5\times$ worse because of the write amplification caused by the nearby shifts. Specifically, with \textit{per-section edge log}, DGAP is able to reduce the write amplification by $6\times$ in the Orkut graph. Additionally, \textit{per-thread undo log} contributes another $13\%$ performance improvement by reducing the high memory allocation and excessive ordering cost of transactions. Finally, placing the \textit{vertex array} in PMs would incur about $2\times$ performance overhead. Placing all the remaining metadata (e.g., PMA tree) in PMs would even double the overhead.

\begin{figure}[ht!]
    \begin{subfigure}[b]{0.48\linewidth}
    	\centering
    	\includegraphics[page=1,width=\linewidth]{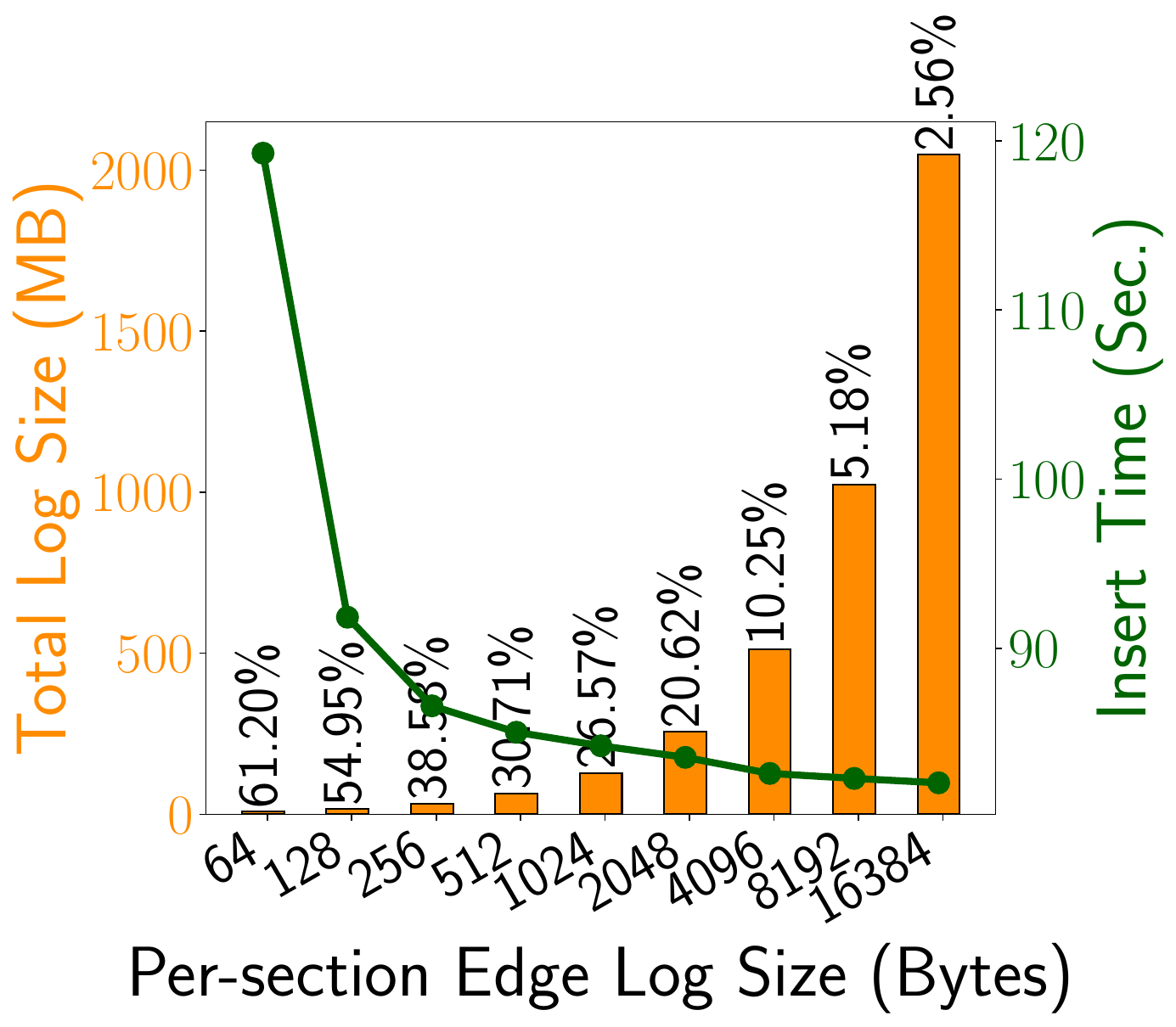}
    	\label{fig:sub-second}
     \vspace{-0.5em}
        \caption{Orkut}
    \end{subfigure}
    \begin{subfigure}[b]{0.48\linewidth}
    	\centering
    	\includegraphics[page=1,width=\linewidth]{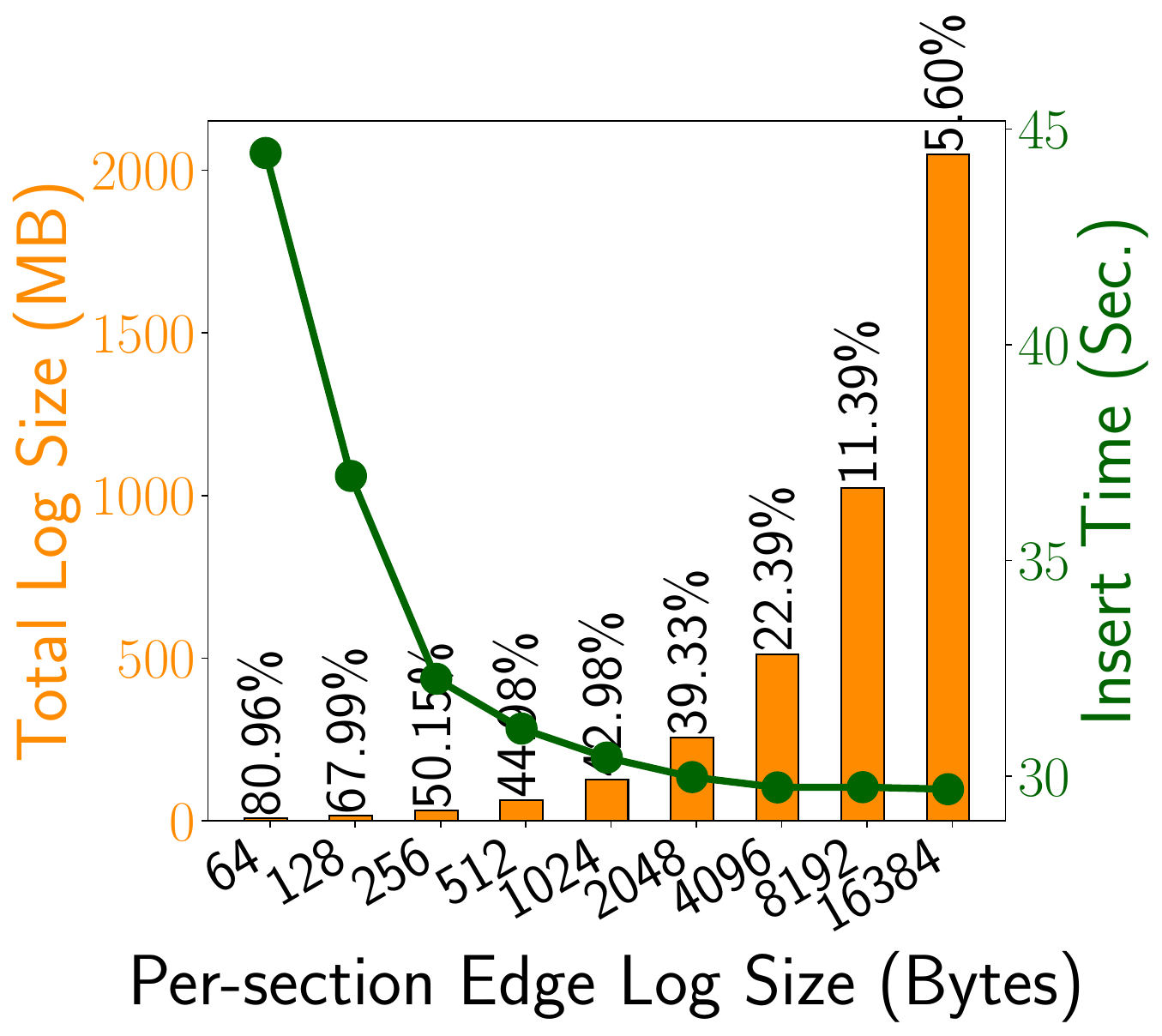}
    	\label{fig:sub-second}
     \vspace{-0.5em}
        \caption{LiveJournal}
    \end{subfigure}
    \caption{\small Impacts of the size of per-section edge log.}
    \label{eval:edge_log_size}
    \vspace{-1em}
\end{figure}

\noindent\textbf{DGAP Configurations Evaluations.}
Besides three system components, DGAP includes a set of configurations, impacting its performance. For example, the size of \textit{per-section edge log} will affect the PM usages as well as the rebalancing frequency, impacting the insertion performance. To evaluate it, we compared how its size, \texttt{ELOG\_SZ}, would impact graph insertion performance and PMs consumption. The results are reported in Fig.~\ref{eval:edge_log_size}. Due to space limits, we only show results for Orkut and LiveJournal graphs. Other graphs have similar patterns. We changed \texttt{ELOG\_SZ} from 64 bytes to 16 KB. The bar length represents the total space needed to store all the \textit{per-section edge log}, which increases proportionally as \texttt{ELOG\_SZ} increase. The labels above each bar further report the percentage-wise utilization of these logs during graph insertions. We can see as the edge log increases, the utilization rate reduces significantly from 80.96\% to 5.60\% as there might not be so many \textit{nearby shifts} to fill the logs. The green line shows the delivered insertion performance based on each log size. It is clear that larger logs reduce the insertion time. But the benefits become much smaller after 2048, which is chosen as default \texttt{ELOG\_SZ} size in DGAP. 

\noindent\textbf{DGAP Recovery Evaluations.}
Each time DGAP reboots, it reloads the metadata into DRAM before operating. Such a normal start is fast. In our evaluation, we found that DGAP spends $1.16$ seconds in rebooting even on the largest Friendster graph.
After crash, DGAP needs to do more housekeeping work to recover system statuses. These steps include scanning the \textit{edge array} and \textit{logs} to recover the inconsistencies caused by the crash. This indicates DGAP crash recovery time will depend on the graph size. However, sequential access in PMs is fast, and so is the DGAP recovery. In our experiment, we found that for the smaller graphs (e.g., Orkut, LiveJournal, and CitPatents), DGAP takes less than $1$ second. For the larger graphs, it may take more than $4$ seconds. But, note that these time costs are for recovery from a crash only.

\section{Related Works}
\label{sec:related}

The works most closely related to ours are NVGRAPH~\cite{lim2019enforcing} and XPGraph~\cite{wang2022xpgraph}. Both frameworks are designed for persistent memory devices. NVGRAPH proposed a dual-version data structure for NVM and DRAM to achieve high-speed data persistence and graph analysis. However, since NVGRAPH was designed before actual persistent memory devices were released, many of its assumptions have later been shown to be inaccurate~\cite{yang2020anempirical}. Consequently, it did not leverage many performance features of PMs. As such, we do not compare DGAP with NVGRAPH, as it wouldn't be a fair comparison. 
Similar to DGAP, XPGraph was designed for and evaluated on Intel Optane PMs, and is essentially a PM-based GraphOne. Through extensive evaluations, we demonstrate that DGAP outperforms XPGraph in both graph updates and graph analysis tasks, highlighting the promising performance of mutable CSR data structures. A recent study~\cite{gill2019single} systematically benchmarks graph processing on PMs. However, this study assumes that persistent memory functions as volatile, larger DRAM serving only graph analysis, which is fundamentally different from DGAP.

In addition to PM-based graph analysis, there has been a large number of PM-based indexing data structures, such as {B+-Tree}~\cite{li2022ssbtree, zhang2022nbtree, chen2020utree, liu2020lbtrees, deukyeon2018endurable, chen2015persistent, ismail2016fptree, jun2015nvtree} and {Hashtable}~\cite{lamar2021pmap, benson2021viper, moohyeon2019write, pengfei2018write, zuo2019level, chen2020lockfree}. Some works~\cite{Lee2019,venkataraman2011consistent,kim2021pactree,haria2020mod,friedman2020nvtraverse,memaripour2020pronto,krishnan2021tips,huang2021ayudante} also proposed general guidelines for porting in-memory data structures to PMs. Many of the DGAP's design choices are aligned with these existing studies, but focus more on graph updates and analysis. 

In addition to PM-based graph frameworks, there are a significant amount of single-node dynamic graph analysis frameworks. We categorize them into in-memory and out-of-core frameworks. 
For in-memory dynamic graph frameworks~\cite{islam2022vcsr, pandey2021terrace, wheatman2018packed, king2016dynamic, firmli2020csr++}, their graphs are not persistent and need rebuilding after a crash or reboot. Even with data periodically synchronized to fast non-volatile storage devices, like PMs, existing in-memory graph frameworks still face the challenges in striking a balance between data loss and graph update speed. Our evaluations on BAL, LLAMA, and GraphOne show naively porting existing in-memory graph frameworks to persistent memory will experience performance issues. DGAP roots from in-memory data structure (mutable CSR) as well, but contains a series of new designs to maximize the performance.
Existing out-of-core dynamic graph frameworks are designed based on slow block-based storage devices~\cite{kyrola2012graphchi,macko2015llama}. 
For example, in LLAMA~\cite{macko2015llama}, newly added edges are first batched up in the delta map and periodically synced to a CSR snapshot. 
Such batch behaviors may not be necessary on persistent memory. 
While, in DGAP, graph changes are immediately visible to analytic tasks.

\section{Conclusion and Future Work}
\label{sec:conclusion}

In this study, we present DGAP, a new graph analysis framework built on persistent memory. DGAP leverages existing DRAM-based mutable Compressed Sparse Row (CSR) graph structure with extensive new designs for persistent memory devices to achieve both efficient graph updates and graph analysis. Our results show DGAP outperforms state-of-the-art dynamic graph frameworks, such as LLAMA, GraphOne, XPGraph on PMs by up to $3.2\times$ in graph updates and $3.77\times$ in graph analysis. Our exploration of DGAP shows that persistent memory is a promising alternative to support efficient dynamic graph analysis. In the future, we plan to further improve DGAP designs, including a Copy-on-Write strategy for Degree Cache and a fine-grained locking mechanism. We also plan to investigate how to extend DGAP to a distributed environment using RDMA in PMs to support even larger graphs. 

\section*{Acknowledgments}
We sincerely thank the anonymous reviewers for their valuable feedback. This work was supported in part by NSF grants {CNS-1852815, CCF-1910727, CCF-1908843, and CNS-2008265}.

\bibliographystyle{ACM-Reference-Format}
{\small
\bibliography{bib}
}

\newpage

\end{document}